\documentclass[epj]{svjour}
\usepackage[dvipdf]{graphicx}
\usepackage{subfig}

\begin{document}

\title{On the relationship between cyclic and hierarchical three-species 
       predator-prey systems and the two-species Lotka--Volterra model}
\titlerunning{Cyclic and hierarchical three-species predator-prey systems}

\author{Qian He \and Uwe C. T\"auber \and R. K. P. Zia}

\institute{Department of Physics, Virginia Tech, Blacksburg, 
           Virginia 24061-0435, U.S.A. \\ 
           \email{heq07@vt.edu; tauber@vt.edu; rkpzia@vt.edu}}


\date{\today}

\abstract{
Stochastic spatial predator-prey competition models represent paradigmatic
systems to understand the emergence of biodiversity and the stability of
ecosystems.
We aim to clarify the relationship and connections between interacting 
three-species models and the classic two-species Lotka--Volterra (LV) model 
that entails predator-prey coexistence with long-lived population oscillations.
To this end, we utilize mean-field theory and Monte Carlo simulations on
two-dimensional square lattices to explore the temporal evolution 
characteristics of two different interacting three-species predator-prey 
systems, namely: 
(1) a cyclic rock--paper--scissors (RPS) model with conserved total particle 
number but strongly asymmetric reaction rates that lets the system evolve 
towards one ``corner'' of configuration space; 
(2) a hierarchical ``food chain'' where an additional intermediate species is
inserted between the predator and prey in the LV model.
For the asymmetric cyclic model variant (1), we demonstrate that the 
evolutionary properties of both minority species in the (quasi-)steady state of
this stochastic spatial three-species ``corner'' RPS model are well 
approximated by the two-species LV system, with its emerging characteristic 
features of localized population clustering, persistent oscillatory dynamics, 
correlated spatio-temporal patterns, and fitness enhancement through quenched 
spatial disorder in the predation rates.
In contrast, we could not identify any regime where the hierarchical 
three-species model (2) would reduce to the two-species LV system.
In the presence of pair exchange processes, the system remains essentially
well-mixed, and we generally find the Monte Carlo simulation results for the 
spatially extended hierarchical model (2) to be consistent with the predictions
from the corresponding mean-field rate equations.
If spreading occurs only through nearest-neighbor hopping, small population 
clusters emerge; yet the requirement of an intermediate species cluster 
obviously disrupts spatio-temporal correlations between predator and prey, and 
correspondingly eliminates many of the intriguing fluctuation phenomena that
characterize the stochastic spatial LV system. 
}

\maketitle

\section{Introduction}
\label{introd}

Spatially extended evolutionary game theory models have found widespread 
applications in population dynamics, with the aim to explore the complex 
spatio-temporal patterns that characterize ecological and biological systems in
which multiple species interact cooperatively or competitively
\cite{MaynardSmith,Hofbauer,Nowak,MayLeonard,May,Maynard,Michod,Sole,Neal}.
Various simplified models have been employed to qualitatively capture and 
quantitatively understand the primary features of interacting multi-species 
systems. 
Two prominent non-trivial paradigmatic representatives are the two-species 
Lotka--Volterra model (LV) 
\cite{Lotka,Volterra,Monetti,Bettelheim,Droz,Antal,Kowalik,McKane,Washenberger,MauroGeorgiev2} 
that describes predator-prey competition and coexistence, and the three-species
rock--paper--scissors (RPS) model 
\cite{MaynardSmith,Hofbauer,Nowak,MayLeonard,Szabo,Sid,FrachKrapBen} 
that encapsulates cyclic competition through predation.
LV and RPS model variants effectively mimic the dynamical evolution of 
coexisting species in interacting multi-species systems or autocatalytic 
chemical reactions \cite{Lotka}, and have been invoked to capture the 
remarkable coexistence of three cyclically competing types of Californian 
lizards \cite{Sinervo,Zamudio}, and of three strains of {\em E. coli} bacteria 
in microbial experiments \cite{Kerr}. 
It was demonstrated that stochastic lattice versions of both LV and RPS models 
may exhibit spatially correlated evolving patterns which turn out to be 
essential in maintaining species coexistence over extended time periods
\cite{Monetti,Antal,Kowalik,Washenberger,MauroGeorgiev2,Sid,FrachKrapBen,Reichenbach1,Reichenbach2,Reichenbach4}.
Ecological systems in nature of course typically contain many more interacting
species; it is therefore vitally important to understand if the intriguing
observed features in two- and three-species models are generic to multi-species
systems as well, and under which circumstances, if at all, an effective 
description in terms of fewer species is appropriate.
In this paper, we consequently aim to explore the relationship between both 
cyclic and hierarchical three-species predator prey systems and the two-species
LV model.
To this end, we shall consider both well-mixed settings that are properly 
described by the corresponding mean-field rate equations, and stochastic 
dynamics on a two-dimensional square lattice with periodic boundary conditions,
where population spreading occurs through nearest-neighbor hopping and/or 
particle exchange.

The classic LV model captures the dynamical evolution of two competing species
of predators and prey; in the well-mixed mean-field limit, the ensuing coupled 
determi\-nistic nonlinear ordinary differential equations yield regular 
periodic population oscillations. 
However, this original non-spatial LV model is often criticized for its 
oversimplification \cite{Neal} and lack of robustness: it is mathematically 
unstable against various model modifications and variations \cite{Murray}. 
Therefore, hopefully more realistic stochastic spatial LV model extensions have
been studied extensively both analytically and computationally; this includes 
the stochastic lattice-based LV model with unrestricted site occupancies 
\cite{Washenberger,Ulrich} and model versions with restricted local carrying
capacity, where only a limited number of individuals may occupy each lattice
site \cite{Monetti,Antal,Kowalik,MauroGeorgiev2,LipowskaD,AFRozenfeld}.
Systems with restricted site occupancy have been shown to display a critical 
extinction threshold for the predator population, where in the thermodynamic
limit a continuous phase transition takes place from an active coexistence 
state to an inactive absorbing state, governed by the power laws of critical
directed percolation 
\cite{Monetti,Antal,Kowalik,MauroGeorgiev2,LipowskaD,AFRozenfeld,Satulovsky,Boccara,Lipowski}.
In the active (quasi-)stationary coexistence state, stochastic spatial LV 
models display essential and robust features such as species clustering and the
emergence of correlated spreading activity fronts 
\cite{Monetti,Antal,Washenberger,MauroGeorgiev2,Panja,Malley,Dunbar,Sherratt,Aguiar}, 
which are associated with the presence of persistent erratic population 
oscillations 
\cite{Monetti,Droz,Antal,Kowalik,Washenberger,MauroGeorgiev2,AFRozenfeld,Lipowski,Provata};
in spatial stochastic systems, these characteristics even pertain when the 
original LV predation reaction is split up into two independent processes 
\cite{MobiliaGeorgiev1}. 
In addition, fitness enhancement for both species is induced by significant 
spatial variability in the predation rates that control the interactions 
between predators and prey \cite{Ulrich}. 

Cyclic competition in three-species RPS systems has been suggested to provide a
robust mechanism to promote species coexistence.
For non-spatial stochastic RPS systems, the temporal evolution always reaches
 one of three extinction states in which two of three species disappear 
\cite{Reichenbach3,Reichen,Berr}. 
In contrast, sufficiently large spatially extended RPS models, particularly 
stochastic two-dimensional RPS systems, are characterized by the coexistence of
all three particle species.
When the cyclic competition reactions do not conserve the total population 
density, complex spatio-temporal structures such as spiral patterns emerge in 
two-dimensional RPS systems 
\cite{Reichenbach1,Reichenbach2,Reichenbach4,Tainaka,Szabo2002,Perc,Tsekouras,QianEPJB}.  
Yet in the simplest spatial stochastic RPS realizations of cyclic three-species
competition, a conservation law for the total number of reactants is 
incorporated, and no spiral patterns become apparent in lattice simulations, 
but instead fluctuating species clusters are observed \cite{Frean,Qian,Matti}.
However, even though spatially correlated clusters emerge in RPS systems with 
conservation law, quenched spatial disorder in the reaction rates does not 
evidently influence the dynamical evolution, provided these rates for the
different species remain comparable \cite{Qian}, which locates the coexistence 
state far away from the ``corners'' of configuration space. 
In this present work, we shall therefore study the effect of quenched spatial 
disorder on RPS systems with strongly {\em asymmetric} reaction rates, whose
coexistence states are shifted closely towards the ``corners'' of configuration
space.

Since both the two-species LV model and cyclic three-species RPS systems have 
attracted considerable attention in the literature, understanding the emergence
of stable biodiversity in population dynamics also requires a better grasp on
potential connections between these two paradigmatic model systems, which might
in turn perhaps enable us to reduce complex interacting multi-species systems 
to effective models with fewer degrees of freedom.
Now let $N$ represent the typical size of the ``living space'' for each species
(e.g., in a lattice-based simulation, $N$ is the typical size of the lattice),
and consider a cyclic three-species RPS model with interaction rates chosen in
such a manner that one species reaches a persistently large population density,
say of $O(1)$, in the (quasi-)steady state (henceforth referred to as the 
majority species), while the other two population densities are reduced to 
$O(1/N)$ (minority species).
Through treating the majority species in the system in effect equivalent to the
empty states in a two-species LV model, the predation processes between the 
majority and the two minority species in such a ``corner'' RPS system can be 
rewritten as the elimination and reproduction processes for the predators and 
prey, respectively, in the LV model.
That is, it should be possible to approximate the evolutionary dynamics of the 
two minority species in the ``corner'' RPS model by the LV model; and the error
limit of this approximation has been shown to be $O(1/N^{2})$ \cite{Zia}.
Another realization of an interacting three-species system is a hierarchical 
``food chain'' model, which is generated by inserting one intermediate species 
between the predator and prey in two-species LV model; the predations between 
predators and prey are then only indirectly generated via their interactions 
with the intermediate population.
It is then an interesting question whether one may still employ the two-species
LV model with fewer degrees of freedom, to effectively approximate the full
stochastic dynamic behavior of the predator and prey populations in such 
spatially extended three-species ``food chain'' systems.

The goal of this work is to explore these two distinct ideas as to how a 
stochastic spatial predator-prey system with originally three interacting 
species might be mapped onto an effective two-species LV model, and thereby 
better comprehend potential mechanisms for complexity reduction in 
multi-species population dynamics.
Our main results can be summarized as follows:

(1) We demonstrate that the two minority species in three-species ``corner''  
RPS system exhibit approximately the same evolutionary dynamics as the 
predators and prey in the two-species LV model: 
We observe quantitatively similar population oscillations and asymptotic 
densities in the (quasi-)steady state, noticeable species clustering, and 
robust fitness enhancement due to spatial variability in the predation rate, 
i.e., larger population densities for both species, associated with reduced 
relaxation times towards coexistence, and more localized species clusters.
These features for the two minority species in the ``corner'' RPS model closely
resemble previous findings for the effect of quenched spatial disorder in the
two-species LV system \cite{Ulrich}. 
We conclude that these numerical results confirm the assertion, based on the
mean-field analysis, that two-species LV models satisfactorily approximate the 
stochastic spatial evolutionary dynamics of the two minority species in the 
three-species ``corner'' RPS system.

(2) We also study the hierarchical three-species ``food chain'' model.
In its spatially well-mixed version, there exists an extinction threshold for
the predator species, just as in the two-species LV model with restricted site 
occupation. 
We find that nearest-neighbor particle pair exchange processes alone are strong
enough to wash out the emergence of species clusters, eliminate the influence 
of quenched spatial disorder and thus promote the appearance of mean-field like
behavior. 
Our numerical simulation results for the coexistence states of these 
effectively well-mixed systems are consistent with the mean-field predictions.
In contrast, if species spreading happens exclusively through nearest-neighbor 
hopping, the spatial ``food chain'' system does not behave according to the 
predictions from the mean-field approximation, and rather distinctive species 
clusters emerge as observed in two-species LV systems.
However, due to the necessity of interspersed clusters of the intermediate 
species, the effective reaction boundaries between the predator and prey 
clusters in these spatially segregated systems are strongly suppressed, and 
correspondingly spatial disorder in the reaction rates still displays only a
minute influence on the evolutionary dynamics of ``food chain'' model.
That is, the predator and prey in the hierarchical three-species ``food chain''
model in a stochastic spatial setting do {\em not} behave akin to their LV
counterparts.

The organization of this paper is as follows: 
In Sec.~\ref{meanftRPS}, we define the lattice-based stochastic 
rock--paper--scissors (RPS) model with conserved total particle number, as well
as the stochastic Lotka--Volterra (LV) model with site occupation number 
restriction, briefly discuss well-established results from mean-field theory, 
and analyze the quantitative relationship between the (quasi-)steady states of 
the ``corner''  RPS model with strongly asymmetric rates and its associated 
two-species LV model. 
In Sec.~\ref{modint}, we explain how we implement our Monte Carlo simulations 
on a two-dimensional lattice with restricted site occupancy, allowing at most a
single individual of either species per site, and introduce the relevant 
physical quantities of interest. 
Then we present and analyze our simulation results for the corner RPS and 
associated LV models, both in the absence and presence of quenched reaction
rate disorder, in Sec.~\ref{modint}. 
In Sec.~\ref{meanftRPSbroken}, we describe the hierarchical three-species 
``food chain'' model and discuss the corresponding analytic results from the
mean-field approach. 
Then the numerical results based on Monte Carlo simulations on a 
two-dimensional lattice with at most one individual per site are presented.
Finally, Sec.~\ref{conclu} concludes with a discussion and interpretation of 
our findings.

\section{Strongly asymmetric ``corner'' RPS model and mean-field analysis}
\label{meanftRPS}

The rock--paper--scissors model (RPS) consists of three zero-sum predator-prey 
interactions \cite{Hofbauer}.
Let $A$, $B$, and $C$ represent the three interacting particle species; the
RPS model is then described by the following binary reactions:
\begin{eqnarray}
  A + B &\to A + A \quad &{\rm with \ rate} \ \lambda \ ;  \nonumber \\
  B + C &\to B + B \quad &{\rm with \ rate} \ \sigma \ ; \label{react1} \\
  C + A &\to C + C \quad &{\rm with \ rate} \ \mu \nonumber 
\end{eqnarray}
(here, we use $\lambda, \sigma, \mu$ to label the corresponding reaction rates 
to resemble the notation used in the two-species LV model 
\cite{Washenberger,MauroGeorgiev2,Ulrich}). 
It is important to note that the total particle population number is always 
conserved by the reactions (\ref{react1}).  
In addition, in our lattice Monte Carlo simulations we shall allow for 
nearest-neighbor exchange processes $X + Y \to Y + X$ with $X, Y \in (A,B,C)$.
We also impose the constraint that at most a single particle (of either 
species) is allowed on each lattice site to mimic limited local carrying 
capacities that result from limited natural resources.
In this present study, the total population density is always set to 1, and the
lattice is hence fully occupied.  

Let $a(t)$, $b(t)$, and $c(t)$ represent the (spatially averaged, see 
Eq.~(\ref{popden}) below) population densities or concentrations of species 
$A$, $B$, and $C$, respectively. 
Within the mean-field approximation, the associated coupled rate equations
\begin{eqnarray}
  \partial_t \, a(t) &=& a(t) \left[ \lambda \, b(t) - \mu \, c(t) \right] \ ,
  \nonumber \\
  \partial_t \, b(t) &=& b(t) \left[ \sigma \, c(t) - \lambda \, a(t) \right] 
  \ , \label{rpsreq} \\
  \partial_t \, c(t) &=& c(t) \left[ \mu \, a(t) - \sigma \, b(t) \right]
  \nonumber
\end{eqnarray}
yield one marginally stable reactive fixed point, describing a three-species 
coexistence state 
$(a^*, b^*, c^*) = \rho (\sigma, \mu, \lambda)/$ $(\sigma + \mu + \lambda)$,
where $\rho = a(t) + b(t) + c(t)$ denotes the conserved total particle density 
\cite{Frean,Qian} (with $\rho = 1$ in this study).  
The three absorbing states $(\rho, 0, 0)$, $(0, \rho, 0)$, and $(0, 0, \rho)$
are all unstable in the mean-field approximation, but one of them will be
eventually reached in any finite stochastic system, after a characteristic
extinction time that increases exponentially with system size 
\cite{Reichenbach1,Reichenbach4,Qian}.
In our simulations, we employ sufficiently large lattices that extinction 
events are extremely unlikely within the run times.
Therefore, systems with comparable reaction rates 
$\lambda \sim \sigma \sim \mu$ approach a (quasi-)steady state far away from 
the ``corners'' of configuration space. 
It has been demonstrated that stochastic fluctuations are comparatively small
in such systems on two-dimensional lattices, and quenched spatial disorder in 
the rates has only minor influence on their dynamical evolution \cite{Qian}. 

However, in the strongly asymmetric limit $\lambda \gg \sigma, \mu$, the 
mean-field stationary densities become to leading order in $1 / \lambda$: 
$(a^*, b^*, c^*) \approx \rho \left( \frac{\sigma}{\lambda}, 
\frac{\mu}{\lambda}, 1 - \frac{\sigma + \mu}{\lambda} \right)$, and the 
corresponding fixed point moves to the vicinity of one of the corners of 
configuration space. 
Due to the relatively large reaction rate $\lambda$, species $C$ acquires a 
very high population density (almost saturated, i.e., $c^* \approx \rho$) in 
the (quasi-)steady ``corner'' state. 
Since the $C$ particles almost uniformly fill the lattice, the minority species
$A$ / $B$ will essentially always encounter a nearest-neighbor partner $C$ to
undergo the third / second reaction in (\ref{react1}).
The evolution of the rare species $A$ and $B$ in the RPS corner state can 
therefore be approximated by the following two-species Lotka--Volterra model 
reactions:
\begin{eqnarray}
  A  \to &\emptyset \qquad &\quad {\rm with \ rate} \ \hat{\mu} = \mu \, c^* 
  \ ; \nonumber \\
  A + B \to &A + A &\quad {\rm with \ rate} \ \lambda \ ; \label{react2} \\
  B \to &B + B &\quad {\rm with \ rate} \ \hat{\sigma} = \sigma \, c^* \ . 
  \nonumber
\end{eqnarray}
Here, the symbol $\emptyset$ denotes the empty state.
Notice that in this stochastic {\em effective} two-species LV model on a 
lattice, empty sites inevitably appear in the system, no matter whether the 
lattice is initially fully occupied or not. 
In effect, the original exchange processes of minority particles $A$ or $B$ 
with the majority $C$ species are replaced with nearest-neighbor hopping to 
empty sites in the resulting two-species LV model. 

In the mean-field approximation, the LV model (\ref{react2}), with the 
constraint that at most one particle is permitted to reside on each site, is 
described by the following two coupled rate equations, where the total density
$\rho$ assumes the role of the overall carrying capacity 
\cite{MauroGeorgiev2,Murray}: 
\begin{eqnarray}
  \partial_t \, a(t) &=& a(t)\left[\lambda \, b(t) - \hat{\mu} \right] \ , 
  \nonumber \\
  \partial_t \, b(t) &=& \hat{\sigma} \, b(t) \left[ 1 - \frac{a(t)+b(t)}{\rho}
  \right] - \lambda \, a(t) b(t) \ . \label{rateq}
\end{eqnarray}
These also follow directly from the RPS rate equations (\ref{rpsreq}) by 
setting $c(t) \approx c^*$; more precisely, in the second rate equation for
$b(t)$, employing the exact conservation relation $c(t) = \rho - a(t) - b(t)$
properly incorporates the limited local carrying capacities, and 
$\hat{\sigma} = \sigma \, \rho$.
The stationary states of the coupled rate equations (\ref{rateq}) consist of 
the linearly unstable extinction state $(a,b) = (0,0)$, another absorbing state
$(0,\rho)$ that is linearly stable for small 
$\lambda < \lambda_c = \hat{\mu} / \rho \approx \mu$, and the coexistence state
$(a_s, b_s) = \left( \frac{\hat{\sigma}}{\hat{\sigma} + \lambda \rho} \left( 
\rho - \frac{\hat{\mu}}{\lambda} \right), \frac{\hat{\mu}}{\lambda} \right)$ 
that exists and becomes linearly stable if $\lambda > \lambda_c$.
Therefore, in the thermodynamic limit the system displays a continuous 
non-equilibrium phase transition from an inactive, absorbing state to the 
active coexistence state at the critical predation rate $\lambda_c$.
The universal power laws that emerge near the extinction threshold are 
characterized by the critical exponents of directed percolation 
\cite{Antal,Kowalik,MauroGeorgiev2,Tauber}.

Linearizing around the active coexistence fixed point leads to
\begin{equation}
  \left( \begin{array}{c} \partial_t \, {\delta a} \\ 
  \partial_t \, {\delta b} \\ \end{array} \right) 
  = L \left( \begin{array}{c} \delta a \\ \delta b \\  \end{array} 
  \right) \ ,
\end{equation}
where $\delta a(t) = a(t)-a_{s}$ and $\delta b(t) = b(t)-b_{s}$, and with the 
linear stability matrix
\begin{equation}
\label{L}
  L = \frac{1}{\lambda \rho (\lambda \rho + \hat{\sigma})} \left( 
  \begin{array}{cc} 
  0 & \ \lambda \rho \hat{\sigma}(\lambda \rho - \hat{\mu}) \\ 
  - \hat{\mu}(\lambda \rho + \hat{\sigma})^{2} & 
  \ - \hat{\sigma}\hat{\mu}(\lambda \rho + \hat{\sigma}) 
  \end{array} \right) \ .
\end{equation}
Its eigenvalues are 
\begin{equation}
  \epsilon_\pm = - \hat{\sigma}\hat{\mu} (2\lambda \rho)^{-1} [1 \pm \sqrt{1 
  - 4 \lambda \rho \hat{\sigma}^{-1} (\lambda \rho \hat{\mu}^{-1} - 1)}] \ .
\label{linevs}
\end{equation}
In the limit $\lambda \gg \sigma, \mu$ in the two-species LV model 
(\ref{react2}), we have $\lambda \rho \hat{\mu}^{-1} = 
\lambda \rho (\mu c^{*})^{-1} \approx \lambda \mu^{-1} \gg 1$ and similarly,  
$\lambda \rho \hat{\sigma}^{-1} \gg 1$, whence the eigenvalues $\epsilon_{\pm}$
for the active coexistence fixed point turn into a complex conjugate pair with 
negative real part.
This demonstrates that, in the strongly asymmetric limit 
$\lambda \gg \sigma, \mu$, the nature of the nontrivial coexistence fixed point
is that of a linearly stable spiral singularity, and the (quasi-)steady state
consequently is approached in an exponentially damped oscillatory manner. 
From the imaginary part of the complex conjugate pair (\ref{linevs}), we 
infer the characteristic oscillation frequency 
$f = \omega / (2 \pi) \approx \sqrt{\hat{\sigma} \hat{\mu}} / (2\pi)$.

Moreover, there are two points worthwhile noticing: \\
(i) The strong asymmetry condition $\lambda \gg \mu, \sigma$ is the 
prerequisite for validating the approximative representation of the stochastic
model (\ref{react1}) through the effective model (\ref{react2}); therefore, in 
the resulting two-species LV system (\ref{react2}) the predation rate $\lambda$
is always larger than the critical threshold 
$\lambda_c = \mu c^* / \rho \ll \lambda$, and correspondingly the system 
resides deep in the active coexistence state. \\
(ii) When implementing the Monte Carlo algorithm for the effective model 
(\ref{react2}), instead of only introducing a local growth limit for the prey 
species $B$ (which in the mean-field limit would be described by the rate
equation $\partial_t \, b(t) = \hat{\sigma} \, b(t) [1 - b(t)/\rho] 
- \lambda \, a(t) b(t)$ \cite{Neal,Washenberger,Murray}), we impose a maximum 
{\em total} population carrying capacity on each site in the simulation. 
Deep in the coexistence state, one would expect either variant to induce 
similar evolutionary dynamics for the model (\ref{react2}). 
First, since $\lambda \gg \sigma$, the stationary mean-field coexistence state 
concentrations for both model variants, namely with growth-limiting constraint 
on either the total or just the prey population, are essentially identical, 
i.e., $(a_s, b_s) \approx \left( \frac{\hat{\sigma}}{\lambda} \left( 1 - 
\frac{\hat{\mu}}{\lambda \rho} \right), \frac{\hat{\mu}}{\lambda} \right)$. 
Second, we shall see that in our stochastic spatial simulations, the 
characteristic oscillation frequencies, which are inferred from the imaginary 
part of the eigenvalues $\epsilon_{\pm}$ in the nontrivial deep coexistence 
state, remain basically unchanged even if we only impose spatial occupancy 
restriction on the prey population \cite{Washenberger,MauroGeorgiev2}.

Altogether, it thus appears reasonable and legitimate on the ground of both
mean-field theory and heuristic considerations that the stochastic spatial
``corner'' three-species RPS system approaches the two-species Lotka--Volterra 
model. 
In the following section, we shall demonstrate through detailed Monte Carlo
simulations that this claim is valid.

\section{Monte Carlo simulation results for the corner RPS and the associated 
         LV models}
\label{modint}

We implement stochastic individual-based Monte Carlo simulations for both the 
``corner'' RPS and the associated LV model on two-dimensional square lattices 
(typically with $N = 256 \times 256$ sites) with periodic boundary conditions 
according to the reaction schemes (\ref{react1}) and (\ref{react2}) that define
these systems. 
The RPS model is moreover subject to nearest-neighbor exchange processes 
$X + Y \to Y + X$, where $X, Y \in (A,B,C)$, while in the LV system we also 
allow nearest-neighbor hopping $X + \emptyset \to \emptyset + X$, in addition 
to particle exchange. 
In either situation, a maximum occupancy of a single particle (of either 
species) is imposed on each lattice site to mimic a finite local carrying 
capacity resulting from limited natural resources. 
At each time step, one individual of any species is chosen at random, and 
simultaneously one of its four nearest neighbors (which might be an empty site 
in the LV system) is selected randomly. 
Subsequently, the center individual and its chosen neighbor undergo the various
reactions defined in the models according to the respective associated rates; 
otherwise, exchange (or hopping for LV) processes take place between the center
particle and its chosen neighbor. 
Once on average all individual particles on the lattice have had a chance to be
selected as center individual, one Monte Carlo step (mcs) is completed; 
therefore, the infinitesimal simulation time step is $\delta t \sim M^{-1}$, 
where $M$ is the total number of individuals on the lattice present at that 
time (a fixed constant only for the RPS system). 
In the following, we are going to investigate four model variants:

1. ``RPS'' --- ``Corner'' rock-paper-scissors model with strongly asymmetric 
rates, $\lambda \gg \sigma, \mu$, whence the system is thus going to eventually
reach one of the corners of configuration space.
In addition, we take the lattice for this model to always be fully occupied, 
$\rho = \rho(0) = 1$; therefore, only exchange processes take place in the 
simulation.

2. ``RPS  with disorder'' --- In order to study the effect of quenched spatial 
disorder, we will treat the predation rate $\lambda$ at each lattice site as a 
random number drawn from a normalized Gaussian distribution truncated at one 
standard deviation on both sides:  
For example, $\lambda \sim {\cal N}(m,n)$ implies that the predation rate 
$\lambda$ is drawn from a truncated normal distribution on the interval 
$[m-n, m+n]$, centered at the value $m$ with standard deviation $n < m$. 
In the simulation, the randomized rates are attached to each site (rather than 
to each individual), and held fixed (quenched) for a given disorder 
realization:
The predation rates $\lambda$ on all sites are determined at the beginning of 
each single Monte Carlo run, and remain unchanged until the next single run is 
initiated. 
Again, the total population density in the simulation is always set to 
$\rho = \rho(0) = 1$ and thus only exchange processes may take place.
All other rates except $\lambda$ stay the same as in the ``RPS'' models.

3. ``LV'' --- Lotka--Volterra model with rates that originate from the rates in
the corresponding corner RPS system: 
If rates $\lambda, \sigma, \mu$ are used in the RPS model, in order to 
appropriately compare the numerical results for both systems, we take 
$\lambda, \hat{\sigma} = \sigma c^*$, $\hat{\mu} = \mu c^*$ for the associated 
LV model, where $c^*$ represents the {\em mean-field} population density for 
the majority species $C$ in the (quasi-)steady state in the asymmetric corner 
RPS model, as determined from the stationary solutions to the rate equations 
(\ref{rpsreq}).
Due to the absence of a conservation law for the total population density, 
empty sites inevitably occur in the course of the simulation runs, whence 
nearest-neighbor hopping is allowed in the simulations in addition to particle 
exchange processes.

4. ``LV with disorder'' --- We treat the predation rate $\lambda$ at each 
lattice as a quenched random variable drawn from the same truncated normalized 
Gaussian distribution as in ``RPS with disorder''. 
Meanwhile, all other rates remain the same as those in the ``LV'' models, and 
again both exchange and hopping processes may take place.

Typically, in order to study the connection between the corner RPS and its 
associated effective LV model, we shall numerically investigate their emerging 
spatio-temporal structures through instantaneous snapshots of the particle 
distributions on the lattice, the temporal evolution of the (spatially 
averaged) population densities, e.g. for species $A$: 
\begin{equation}
  a(t) = \langle n_A(j, t) \rangle = \frac{1}{N} \sum_{j} n_A(j, t) \ , 
\label{popden}
\end{equation}
the temporal Fourier transform of these signals,
\begin{equation}
  a(f) = \int a(t) \, e^{2 \pi i f t} \, dt \ ,
\label{denftr}
\end{equation}
and the equal-time two-point correlation functions in the (quasi-)steady state,
for example
\begin{equation}
  C_{AB}(x,t) = \langle n_A(j+x,t) \, n_B(j,t) \rangle - a(t) \, b(t) \ ,
\label{quant}
\end{equation}
where $j$ and $j+x$ represent square lattice site indices.

\subsection{Self-organization in the coexistence phase of the two minority 
            species in the corner RPS model}

\begin{figure*}[!t]
\label{Fig1}
\begin{center}
\subfloat[Minority population densities.]{\label{fig1:density}
\includegraphics[width=0.50\textwidth]{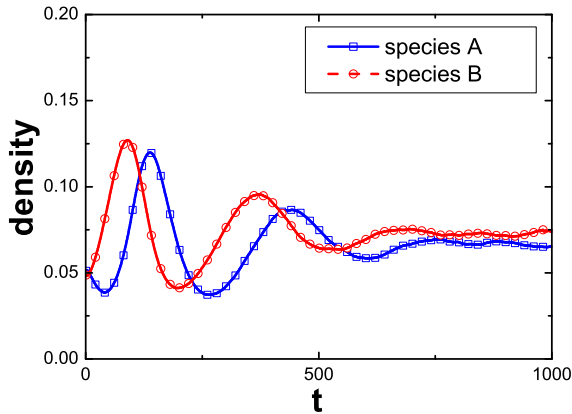}} 
\subfloat[Lattice snapshot at $t = 1000$ mcs.]{\label{fig1:snpsh}
\includegraphics[width=0.35\textwidth]{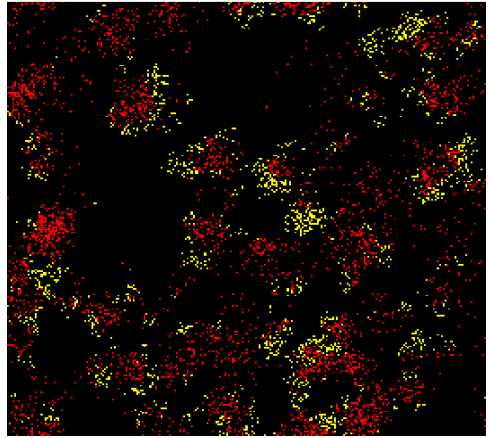}} 
\end{center}
\caption{{\it (Color online.)} (a) Temporal evolution for the population 
  densities of the minority species $A$ and $B$ for an asymmetric RPS system 
  with rates $\lambda = 0.8$ and $\sigma = \mu = 0.07$ on a square lattice with
  $N = 256 \times 256$ sites, periodic boundary conditions, and initial 
  densities $a(0) = b(0) = 0.05$, $c(0) = 0.9$; the data are averaged over 50 
  simulation runs.
  (b) Corresponding snapshot of the spatial particle distribution at $t = 1000$
  mcs (yellow/light gray: $A$, red/gray: $B$, black: $C$); note that there is 
  no empty site in the lattice, since $\rho = a(0) + b(0) + c(0) = 1$ is 
  conserved under the temporal evolution.}
\end{figure*}
In Fig.~\ref{fig1:density}, the population densities for the minority species 
$A$ and $B$ in an asymmetric ``corner'' RPS system model on a square lattice 
with rates $\lambda = 0.8$ and $\sigma = \mu = 0.07$ are plotted as functions 
of simulation time $t$ (in mcs).  
While the population densities for species $A$ and $B$ as obtained from the 
mean-field approximation for the RPS model would be identical,
$\frac{\rho}{\lambda + \sigma + \mu}(\sigma, \ \mu) \approx (0.074, \ 0.074)$, 
the corresponding population densities in the associated LV model assume 
unequal values, $(a_s, b_s) = (0.065, 0.075)$ in the (quasi-)steady state.
The LV predictions are remarkably close to the measured values 
$(0.064 \pm 0.003, 0.074 \pm 0.002)$ (averaged over 50 runs at $t = 1000$ mcs, 
see Fig.~\ref{fig1:density}), clearly distinct for species $A$ and $B$. 
In particular, the persistent unequal population densities (starting from 
$t \approx 600$ mcs in Fig.~\ref{fig1:density}) support the population density 
inequality for minority species $A$ and $B$ in the (quasi-)steady state as
statistically reliable. 
At the beginning of the simulations, we observe marked population oscillations,
see Fig.~\ref{fig1:density}, reflecting the formation of complex 
spatio-temporal structures, i.e., expanding activity fronts as familiar in the
stochastic spatial LV system, noticeable as minority speci\-es clusters in 
Fig.~\ref{fig1:snpsh}. 
The emergence of such localized species clusters promotes species coexistence, 
reduces the oscillation amplitudes in the system, and eventually allows the 
system to evolve into the (quasi-)steady coexistence state. 
In light of these observations, we conclude that the temporal evolution of the
two minority species $A$ and $B$ in the stochastic spatial corner RPS model 
with strongly asymmetric rates can indeed be faithfully approximated by the 
predator-prey behavior in the related two-species LV model.
We shall further support this assertion through additional observations in the 
following subsection.

\subsection{Fitness enhancement of the minority species due to spatial 
            variability in the corner RPS model}

\begin{figure*}[!t]
\begin{center}
\subfloat[Temporal evolution $a(t)$.]{\label{fig2:density}
\includegraphics[width=0.43\textwidth]{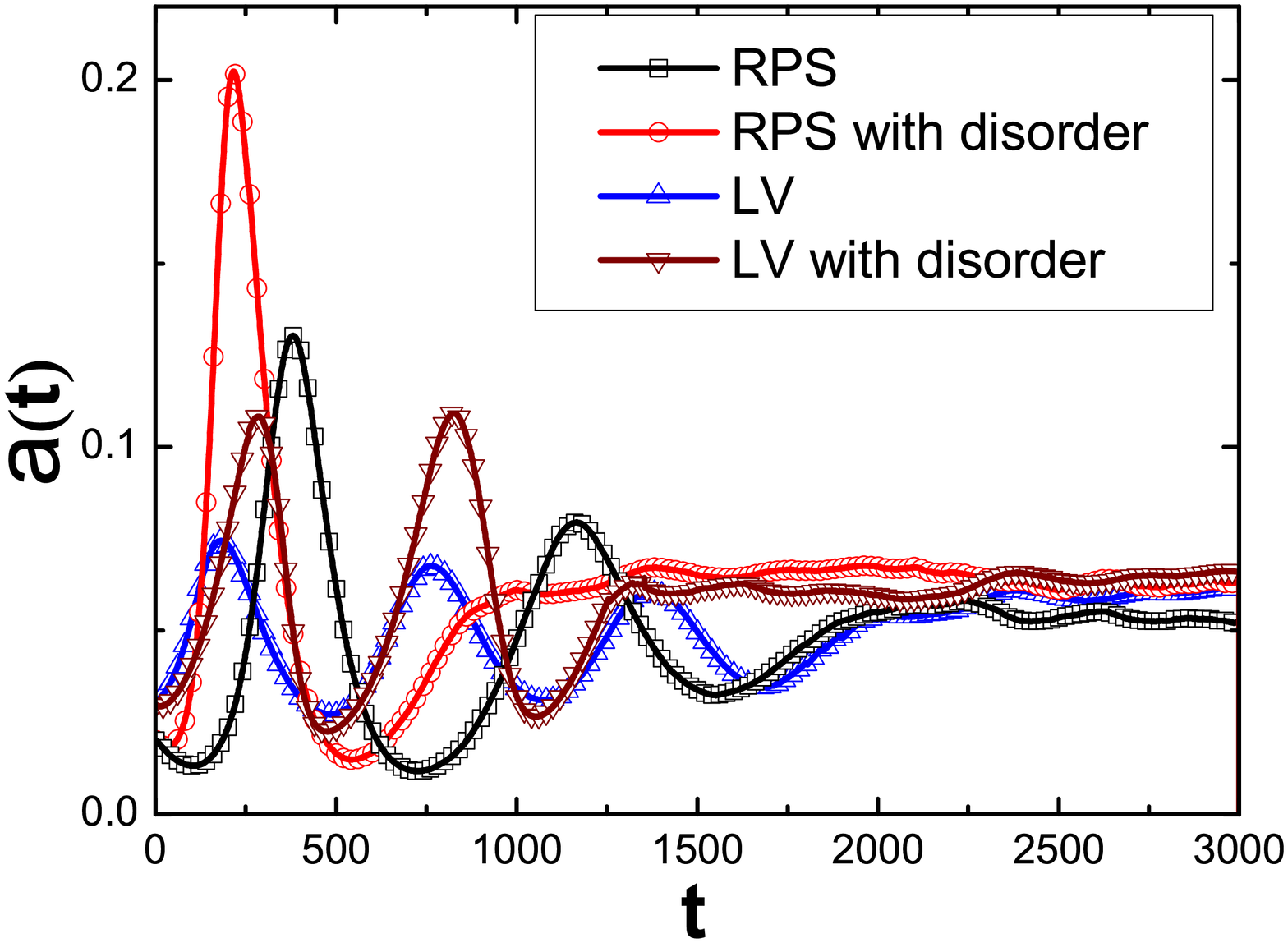}}
\subfloat[Fourier transform $|a(f)|$.]{\label{fig2:FFT}
\includegraphics[width=0.43\textwidth]{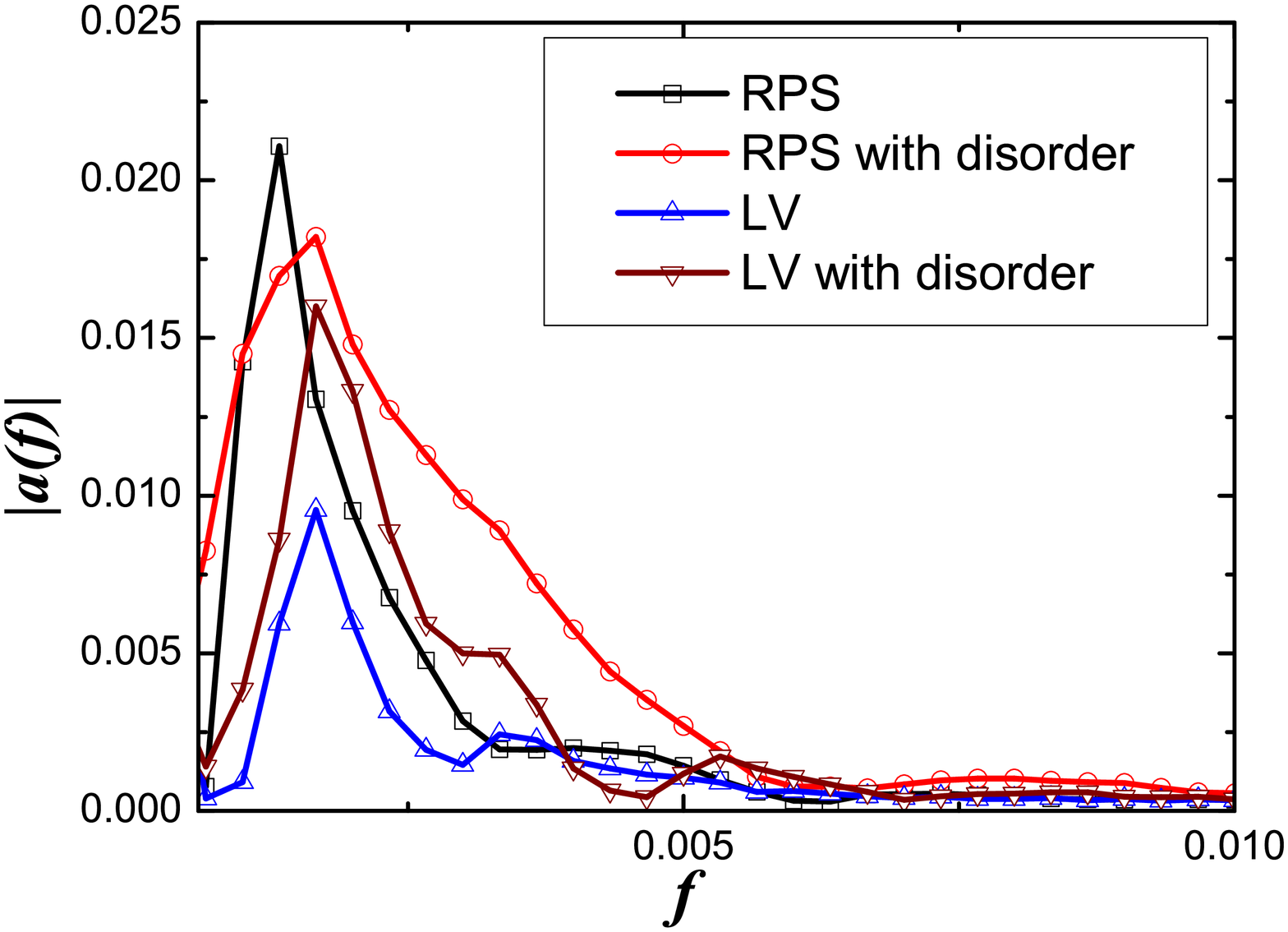}} \\
\subfloat[Correlation function $C_{AA}(x)$.]{\label{fig2:Caa}
\includegraphics[width=0.43\textwidth]{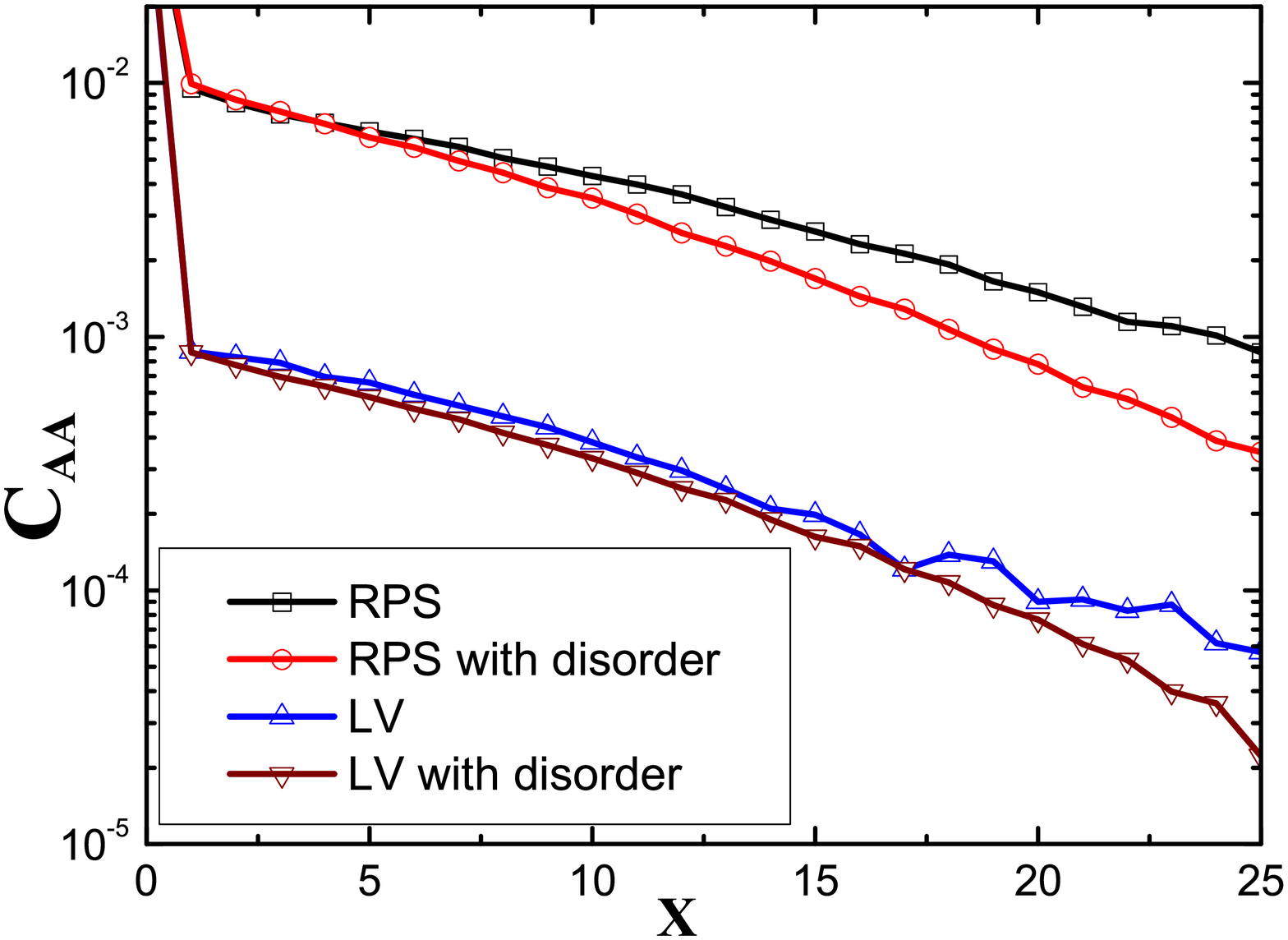}}
\subfloat[Cross-correlation $C_{AB}(x)$.]{\label{fig2:Cab}
\includegraphics[width=0.455\textwidth]{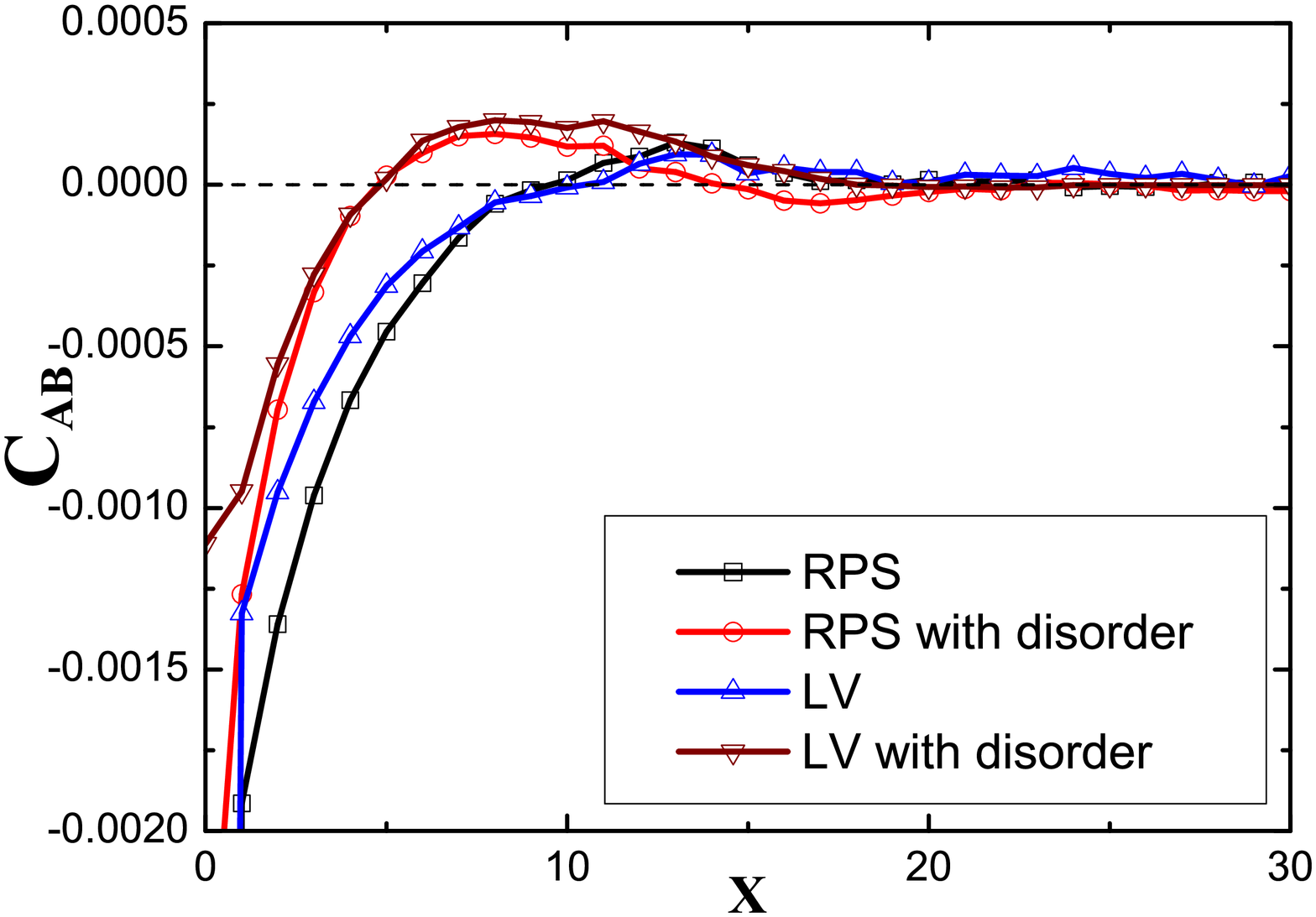}}
\end{center}
\caption{{\it (Color online.)} Quantitative observables for a stochastic corner
   RPS model with rates $\lambda = 0.5$, $\sigma = \mu = 0.03$, and initial
   densities $a(0) = b(0) = 0.02$, $c(0) = 0.96$, and its corresponding LV 
   system with $\lambda = 0.5$, $\hat{\sigma} = \hat{\mu} = 0.027$ and 
   $a(0) = b(0) = 0.03$ from simulation runs on a square lattice with 
   $N = 256 \times 256$ sites (periodic boundary conditions).
   In the presence of spatial disorder, $\lambda \sim {\cal N}(0.5, 0.4)$. 
   All simulation results were averaged over 50 runs, and the correlation 
   functions in (c) and (d) were measured at $t = 3000$ mcs.}
\label{Fig2}
\end{figure*}
In order to further and quantitatively characterize the connection between the
strongly asymmetric corner RPS system and its corresponding effective LV model,
we next study the influence of quenched spatial disorder on the evolution of 
either system. 
Since spatial variability turns out to have very similar effects on species $A$
and $B$, we mainly show the results associated with species $A$; a complete 
listing of data for characteristic observables is provided in 
Table~\ref{qlist}.
We depict the temporal evolution of the population density 
$a(t)= \langle n_A(j,t) \rangle$, its Fourier transform $a(f)$, and the spatial
auto- and cross-correlation functions $C_{AA}$ and $C_{AB}$, respectively, in 
the (quasi-)steady state (here, obtained at $t = 3000$ mcs) in Fig.~\ref{Fig2}.
The default rates are set to $\lambda = 0.5$, $\sigma = \mu = 0.03$ in the 
corner RPS model, and correspondingly $\lambda = 0.5$, 
$\hat{\sigma} = \hat{\mu} = 0.027$ in the associated LV model, and 
$\lambda \sim {\cal N}(0.5, \ 0.4)$ in both corner RPS and LV systems with 
quenched spatial disorder in the rate $\lambda$.
Note that within the mean-field approximation, the population density for 
species $A$ in the (quasi-) steady state is $a_s \approx 0.048$, and the 
characteristic critical predation rate threshold is 
$\lambda_c = 0.027 < \lambda$, guaranteeing that the associated LV system is 
located in the active (quasi-)steady coexistence region of the phase diagram; 
it is also worth noticing that the minimum predation rate in the model variants
with quenched spatial disorder is $0.1$, which is still well above the critical
threshold, eliminating the possibility of any species extinction in our large
lattices, even in the presence of spatial rate variability. 

Figure~\ref{fig2:density} shows the simulated average population densities as 
function of time for the minority species $A$ in both corner RPS and associated
LV systems, in the absence and presence of quenched spatial disorder.
In the stationary state, the $A$ species concentrations (averaged over 50 runs 
at $t = 3000$ mcs) become $0.057$ for the corner RPS model, $0.060$ for the
associated LV model, $0.063$ for the RPS model with quenched spatial disorder, 
and $0.068$ for LV model with spatial disorder (see Table~\ref{qlist}). 
Thus, within the error bars, the population density ($0.057$) for species $A$ 
in the ``pure'' corner RPS model coincides with the simulation result ($0.060$)
for the associated stochastic spatial LV model.
Moreover, we observe that spatial variability in the corner RPS model enhances 
the fitness of the minority species, as measured by the asymptotic stationary
concentrations, by $\sim 11\%$. 
This fitness enhancement is in qualitative accord with earlier investigations
on the effect of quenched randomness in the predation rate for the the 
two-species LV model without site restriction \cite{Ulrich}, but turns out to 
be quantitatively less significant than the $\sim 25\%$ found there; it is also
slightly smaller than the $\sim 13\%$ enhancement obtained directly in our
simulations of the two-species LV model with restricted site occupation number,
see Fig.~\ref{fig2:density} and Table~\ref{qlist}.

We may attribute these differences to the fact that spatial disorder also 
affects the stationary population density $c^*$ of the majority species $C$ in 
the (quasi-)steady state of the corner RPS model, which in turn implies locally
varying effective rates $\hat{\sigma}$ and $\hat{\mu}$ for the associated 
two-species LV model.
This becomes apparent already on the mean-field description level, where 
$c^* = \frac{\lambda}{\lambda + \sigma + \mu}$ and 
$\hat{\sigma} = \sigma c^*$, $\hat{\mu} = \mu c^*$. 
Then for the minority species concentration deep in the coexistence state of 
the corner RPS model one obtains $a_s \approx \frac{\hat{\sigma}}{\lambda} 
= \frac{\sigma}{\lambda + \sigma + \mu}$.
Similarly, for the two-species LV model with site restrictions,
$a_s \approx \frac{\sigma}{\lambda + \sigma}$, whereas 
$a_s = \frac{\sigma}{\lambda}$ in the LV model with infinite local carrying 
capacity.
In the latter case, a broad distribution of predation rates $\lambda$ strongly
biases the system towards small values that considerably enhance the associated
(quasi-)steady state density of $A$ species \cite{Ulrich}.
For the RPS system or the LV model subject to site occupation restrictions, the
additional rates in the denominators reduce the significance of this 
low-$\lambda$ bias and ensuing concentration enhancement, and consequently the
overall fitness enhancement is less pronounced.

As one would expect, the emergence of species clusters results in remarkable 
population oscillations at the beginning of the simulation runs, as seen in 
Fig.~\ref{fig2:density}. 
However, as shown in Fig.~\ref{fig2:density}, the transient period for the 
system evolving from the initial to the eventual active (quasi-)steady state is
quite distinct for the different models.
The corresponding Fourier-transformed density signal, which reflects both the 
characteristic oscillation frequency in the steady state and the associated 
characteristic relaxation time, is displayed in Fig.~\ref{fig2:FFT}.
The Monte Carlo simulations exhibit the same peak frequency in the (quasi-) 
steady state for all four models, $f \approx 0.0014$; this characteristic peak 
frequency in the stochastic lattice systems is markedly reduced by $\sim 60\%$ 
as compared with the (linearized) mean-field prediction 
$f \approx \sqrt{\hat{\mu}\hat{\sigma}}/(2\pi) \approx 0.0043$. 
This remarkable decrease in the typical population oscillation frequency can be
attributed to the renormalization by stochastic fluctuations in the spatial 
system \cite{Washenberger,MauroGeorgiev2,Tauber}. 
Moreover, the relaxation time $\tau = 1/\Gamma$, where $\Gamma$ denotes the 
full width at half maximum of the oscillation peak in Fourier space, 
characterizes the typical time scale for the system to relax towards the 
(quasi-)steady state, and thus represents a measure of stability for these 
systems against external perturbations. 
Significantly, as demonstrated by the data in Fig.~\ref{fig2:FFT} and 
Table~\ref{qlist}, the relaxation time in the RPS model with quenched spatial 
disorder is reduced by $\sim 50\%$ to $\sim 800$ mcs as compared with 
$\tau \sim 1800$ mcs in the corner RPS model without spatial rate variability.
\begin{table*}[!t]
\begin{center}
\begin{tabular}{|l|l|l|l|l|l|l|l|} \hline
  & $a_{s}$ & $b_{s}$ & $\tau_{A/B}$ & $l_{AA}$ & $l_{BB}$ & $l_{AB}$ \\ \hline
  RPS model & 0.057 $\pm$ 0.003 & 0.060 $\pm$ 0.002 & $\sim$ 1800 mcs & 
  13.2 $\pm$ 0.6 & 16.0 $\pm$ 2.0 & $\sim$ 13.0\\ \hline
  RPS with disorder & 0.063 $\pm$ 0.002 & 0.067 $\pm$ 0.003 & $\sim$ 800 mcs & 
  9.4  $\pm$ 0.2 & 13.0 $\pm$ 1.0 & $\sim$ 8.0 \\ \hline
  LV model & 0.060 $\pm$ 0.003 & 0.063 $\pm$ 0.003 & $\sim$ 2000 mcs & 
  12.8 $\pm$ 0.5 & 15.5 $\pm$ 0.5 & $\sim$ 13.0 \\ \hline
  LV with disorder & 0.068 $\pm$ 0.002 & 0.070 $\pm$ 0.003 & $\sim$ 1300 mcs & 
  10.3 $\pm$ 0.3 & 11.2 $\pm$ 0.2 & $\sim$ 9.0 \\ \hline
\end{tabular}
\caption{Characteristic physical quantities measured by Monte Carlo simulations
   for the four stochastic spatial RPS and LV models whose features are 
   displayed in Fig.~\ref{Fig2}. 
   The population densities ($a_s, b_s$) in the (quasi-)steady state are 
   obtained as the averages over 50 simulation runs at $t = 3000$ mcs (see 
   Fig.~\ref{fig2:density}), the associated errors are measured as the standard
   deviations of the 50 data; and the relaxation time $\tau$ is found as the 
   inverse of the full width at half maximum of the frequency peak in 
   Fig.~\ref{fig2:FFT}.
   The correlation lengths $l_{AA}$ and $l_{BB}$ are extracted by fitting the 
   simulation data for $C_{AA}$ and $C_{BB}$ to exponential functions (see 
   Fig.~\ref{fig2:Caa}); and the separation length $l_{AB}$ is measured as the 
   site location with the maximum value for $C_{AB}$ (see Fig.~\ref{fig2:Cab}).}
\label{qlist} 
\end{center}
\end{table*}

The enhancement of both predator and prey fitness resulting from spatial 
variability in the stochastic lattice LV model is based on the emergence of 
more localized correlated spatio-temporal structures \cite{Ulrich}. 
In all four model systems investigated here, such species clustering appears as
well, see Fig.~\ref{fig1:snpsh}. 
Compared to the RPS/LV model without spatial disorder, the corresponding RPS/LV
system with spatial variability in the reaction rate $\lambda$ displays larger 
initial oscillation amplitudes (c.f. Fig.~\ref{fig2:density}), implying that 
spatial disorder in the predation rate tends to gather species closer, 
resulting in more localized and thus enhanced population growth spurts.   
To better understand the underlying spatial structures in our systems, we 
depict the static two-point auto-correlation functions in Fig.~\ref{fig2:Caa} 
and cross-correlation functions in the (quasi-) steady state in 
Fig.~\ref{fig2:Cab}.  
From these plots, we extract the typical correlation length $l_{AA}$, which 
measures the spatial extent of species $A$ clusters by fitting the simulation 
data $C_{AA}$ with an exponential function $\exp(-|x|/l_{AA})$, and obtain the 
predator-prey separation length $l_{AB}$, defined as the location where 
$C_{AB}$ assumes its maximum positive value. 
In the corner RPS model, the correlation and separation lengths turn out to be 
$l_{AA} = 13.2$ (in units of lattice spacing) and $l_{AB} = 13.0$, 
respectively, both of which coincide within the statistical error bars with the
simulation results for the associated LV model, for which we find 
$l_{AA} = 12.8$ and $l_{AB} = 13.0$, as listed in Table~\ref{qlist}. 
However, in the corresponding disordered RPS system, we measure markedly 
reduced values for the correlation and separation lengths, namely $9.4$ and 
$8.0$. 
This closer clustering near sites with locally favorable rates permits an
overall larger number of population centers, and consequently enhances the net
population densities for the minority species in the system with spatial
variability in the rate $\lambda$.

In summary, as suggested on the basis of mean-field considerations, we find 
that the minority species $A$ and $B$ in our spatially extended stochastic RPS 
model with strongly asymmetric rates behave like the predator and prey in the
coexistence phase of the two-species Lotka--Volterra model. 
Indeed, similar to our earlier findings for the lattice LV model \cite{Ulrich},
we observe that quenched spatial disorder can markedly enhance the fitness of
both minority species. 
This feature distinguishes the corner RPS model from the perhaps more common 
RPS system with comparable reaction rates, whose stationary state resides close
to the center of configuration space; in that situation, quenched spatial 
reaction rate disorder hardly influences the dynamical evolution of either 
species \cite{Qian,QianEPJB}.

\section{Hierarchical three-species ``food chain'' model}
\label{meanftRPSbroken}

In Section~\ref{modint}, we have demonstrated how the properties of the
coexistence state in the two-species Lotka--Volterra model emerge as the
rock--paper--scissors system is moved towards one of the corners of 
configuration space by choosing strongly asymmetric rates.
Another natural idea to generate the LV model is to consider a hierarchical 
three-species ``food chain'' system, where an intermediate species $C$ is 
inserted between the predators $A$ and prey $B$.
The question then becomes: will a stochastic hierarchical three-species food 
chain model on a lattice in a certain limit again recover the properties of the
spatial two-species LV system?
To this end, we investigate the following coupled stochastic reactions that 
define our three-species food chain:
\begin{eqnarray}
  A      \to& \emptyset \qquad& \quad {\rm with \ rate} \ \mu \ ; \nonumber \\
  A + C  \to& A + A& \quad {\rm with \ rate} \ \lambda \ ;  \nonumber \\
  C + B  \to& C + C& \quad {\rm with \ rate} \ \lambda \ ; \label{react3} \\
  B      \to& B + B& \quad {\rm with \ rate} \ \sigma \ . \nonumber 
\end{eqnarray}
Species $A$ and $B$ in this model behave as predators and prey, respectively,
while the intermediate population $C$ preys on species $B$ and is preyed upon 
by species $A$. 
In order to perhaps closely approach the two-species LV system and mainly study
the dynamic behavior of species $A$ and $B$, we here use identical reaction 
rates $\lambda$ for the $A$-$C$ and $B$-$C$ predation processes.
We remark that upon identifying the empty state with a fourth species, this
system can be mapped onto a cyclic four-species predator-prey model, see, e.g.,
Refs.~\cite{Durney,Dobrinevski}, with a peculiar choice of rates.
In our Monte Carlo simulations for a spatially extended system on a square 
lattice with periodic boundary conditions, we allow at most one particle per 
site. 
The resulting mean-field rate equations now read:  
\begin{eqnarray}
\label{RE}
  \partial_t \, a(t) &=& a(t) \left[ \lambda \, c(t) - \mu \right] \ , 
  \nonumber \\
  \partial_t \, c(t) &=& \lambda \, c(t) \left[ b(t) - a(t) \right] \ , 
\label{rateq2} \\
  \partial_t \, b(t) &=& \sigma \, b(t) \left[ 1 - a(t) - b(t) - c(t) \right] 
  - \lambda \, b(t) c(t) \ , \nonumber 
\end{eqnarray}
since only the reproduction process for the species $B$ requires the presence 
of an empty site in its immediate neighborhood, whereas newly generated $A$ and
$C$ particles just replace their respective prey.
As stationary solutions of the coupled rate equations (\ref{rateq2}), one finds
one linearly unstable absorbing point $(a, c, b) = (0, 0, 1)$, an entire 
absorbing line $(0, c, 0)$, where $c \in [0, 1]$, and one active coexistence 
fixed point:
\begin{eqnarray}
  (a_s, c_s, b_s) = \left( \frac{\lambda \sigma - \mu (\lambda + \sigma)}
  {2 \lambda \sigma} \ , \frac{\mu}{\lambda} \ ,
  \frac{\lambda \sigma - \mu (\lambda + \sigma)}{2 \lambda \sigma} \right) . 
  \quad
\end{eqnarray}
As in the two-species LV model with site occupation restrictions, there exists 
a critical threshold for the predation rate 
$\lambda_c = \frac{\mu \sigma}{\sigma - \mu}$ in this hierarchical food chain 
model, where in the thermodynamic limit a phase transition occurs from the 
inactive absorbing states to the active coexistence state: 
When the predation rate is below the threshold ($\lambda < \lambda_c$), the 
system evolves towards extinction for the $A$ and $B$ population $(0, c, 0)$; 
otherwise, the system eventually reaches the three-species coexistence state 
$(a_s, c_s, b_s)$. 
However, in stark contrast with the mean-field analysis for the two-species LV 
model with site restrictions, the population densities for the predators $A$ 
and the prey $B$ in this food chain system are always the same in the 
stationary coexistence state, in order for the intermediate species $C$ to 
possess a nonzero stationary population density.  
Other important distinct spatio-temporal properties for the spatially extended 
hierarchical food chain model will be discussed in the following two 
subsections.

\subsection{Monte Carlo simulations: mean-field like behavior}
\label{meanfieldbehavior}

\begin{figure*}[!t]
\label{Fig3}
\begin{center}
\subfloat[Normalized evolution trajectories]{\label{fig3:triangular}
\includegraphics[width=0.43\textwidth]{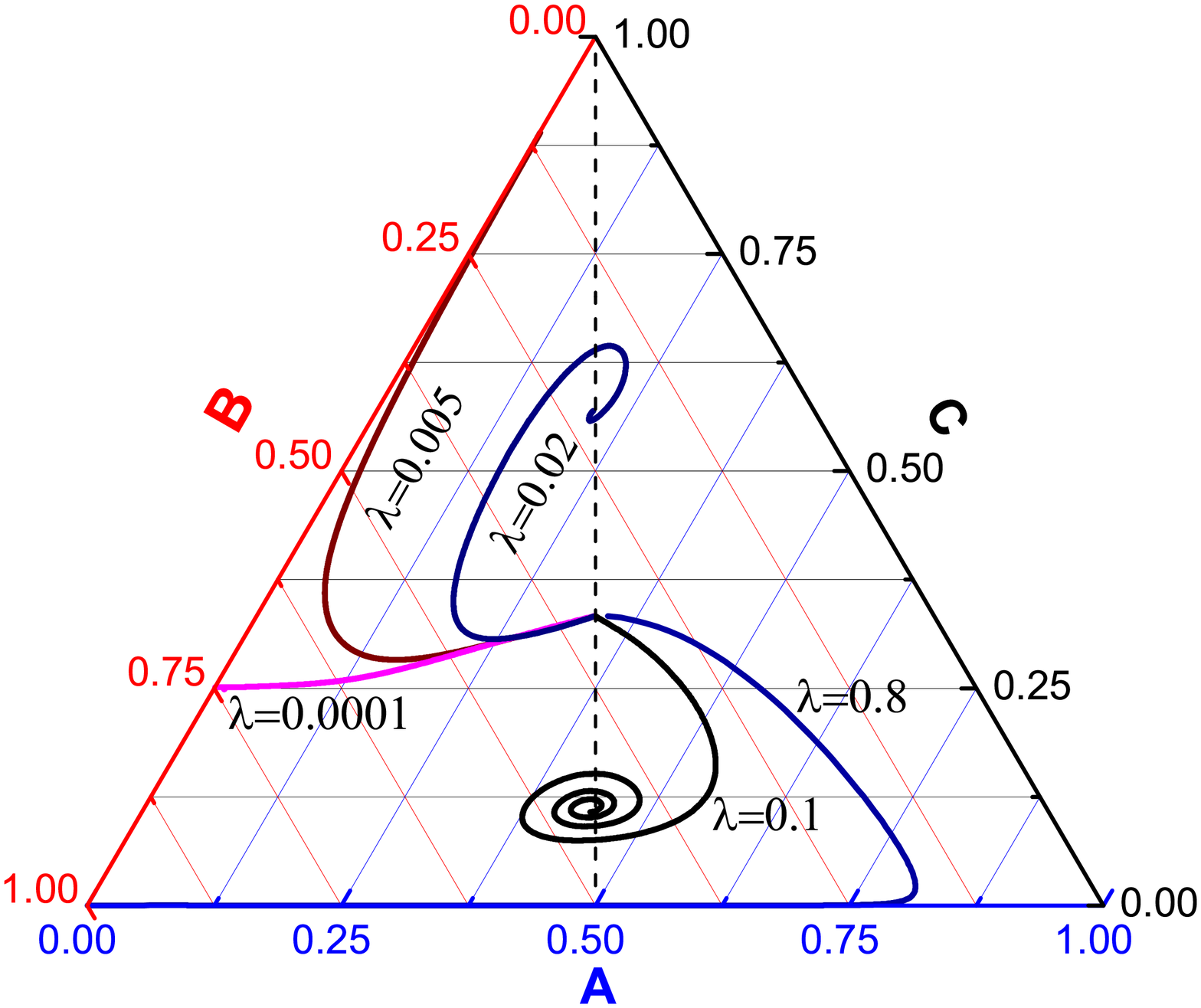}}
\subfloat[$c_s \sim 1/\lambda$ in the (quasi-)steady state]{\label{fig3:lambda}
\includegraphics[width=0.43\textwidth]{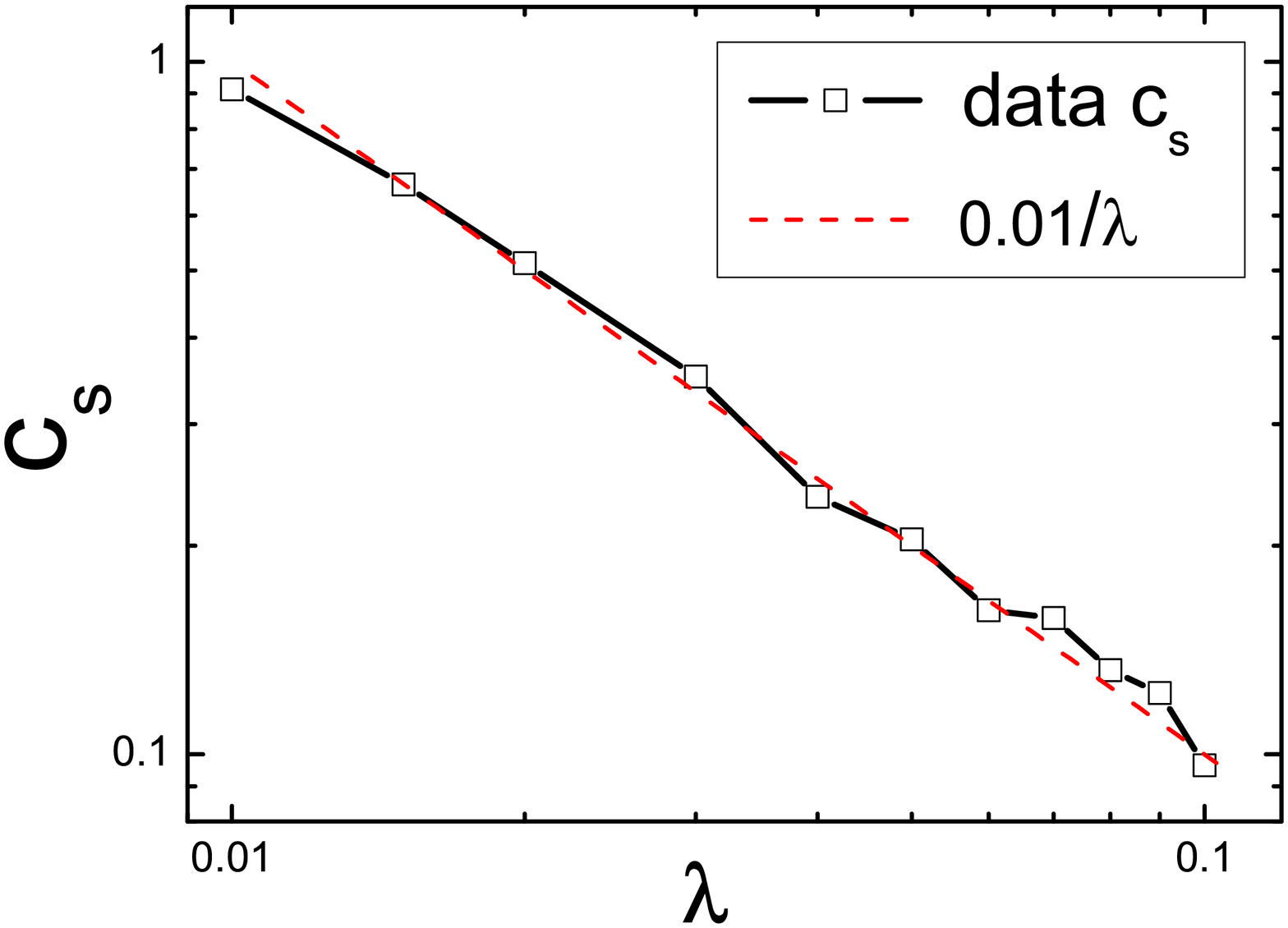}} \\
\hspace{-0.7in}\subfloat[Population densities for 
  $\lambda = 0.1$]{\label{fig3:density}
\includegraphics[width=0.48\textwidth]{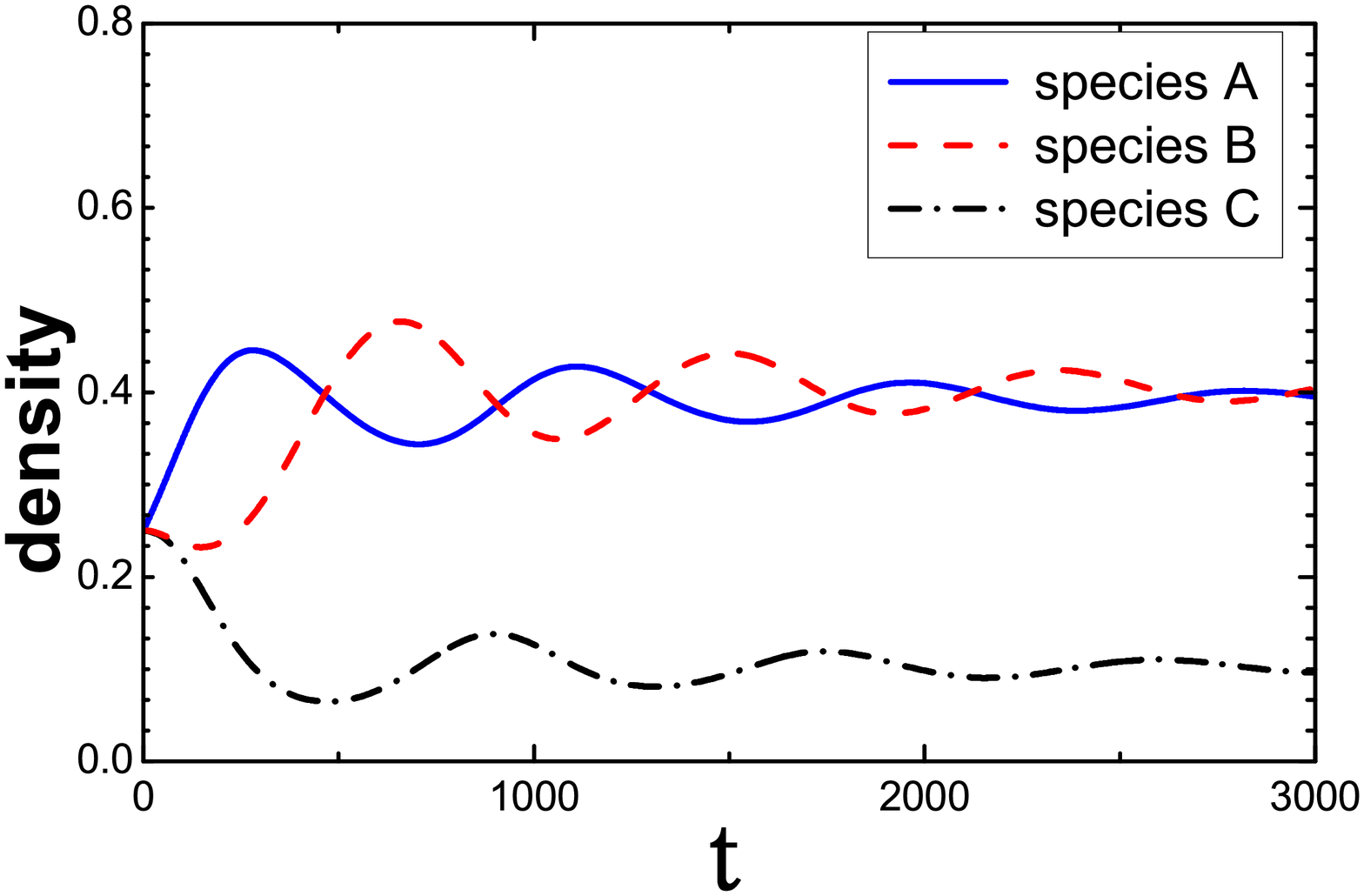}}
\hspace{0.5in}\subfloat[Snapshot at $t = 3000 \ mcs$ for 
  $\lambda = 0.1$]{\label{fig3:snpsh}
\includegraphics[width=0.33\textwidth]{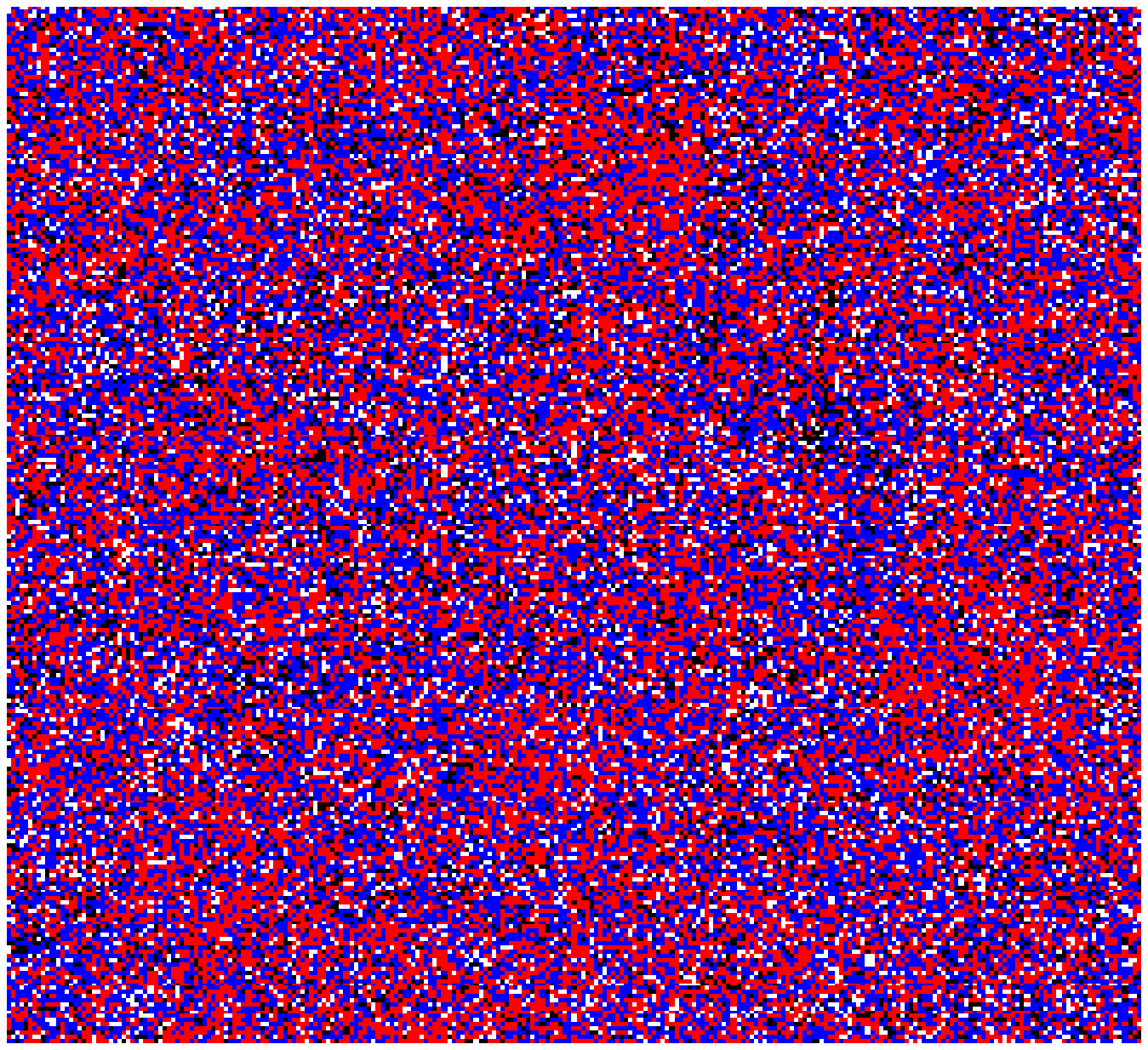}}
\end{center}
\caption{{\it (Color online.)} Observables for a stochastic three-species
   hierarchical food chain system in two dimensions with $N = 256 \times 256$ 
   sites, reaction rates $\mu = 0.01, \sigma = 0.1$, and $\lambda$ as 
   indicated; the simulations started with equal initial densities 
   $a(0) = b(0) = c(0) = 1/3$ in (a) and (b), and $a(0) = b(0) = c(0) = 0.25$ 
   in (c) and (d); the data were averaged over 50 simulation runs; color coding
   in (d): blue/dark gray: $A$, red/gray: $B$,  black: $C$, white: empty.}
\end{figure*}
In the Monte Carlo simulation runs, we first permit both nearest-neighbor 
particle exchange and hopping processes to generate population spreading.
In Fig.~\ref{fig3:triangular}, we depict the typical normalized trajectories 
for the population densities in systems with varying predation rates $\lambda$.
The other rates $\mu = 0.01$, $\sigma = 0.1$ are unchanged, whence the 
critical predation rate threshold value remains fixed at 
$\lambda_c = \frac{\mu \sigma}{\sigma - \mu} \approx 0.011$. 
When the predation rate is below this threshold, the trajectories 
($\lambda = 0.0001, 0.005$ in Fig.~\ref{fig3:triangular}) reach the boundaries 
in the phase space spanned by the the population densities, which constitute
inactive absorbing states. 
Upon increasing the predation rate ($\lambda = 0.02, 0.1$ in 
Fig.~\ref{fig3:triangular}), we observe that the system evolves to the active 
coexistence state. 
While for small predation rates just above the threshold ($\lambda = 0.02$ in 
Fig.~\ref{fig3:triangular}) the fixed point is a node, it changes character and
becomes a spiral focus for larger values, as evidenced by the spiral 
trajectory in the phase graph ($\lambda = 0.1$ in Fig.~\ref{fig3:triangular}). 
This fixed point evolution as function of the predation rate is remarkably 
similar to the situation in two-species LV model with site restriction 
\cite{MauroGeorgiev2}. 
As discussed earlier, in the active coexistence state the predator $A$ and prey
$B$ species should possess the same population densities in order to let the 
intermediate species $C$ also have nonzero stationary population density
(see the dashed line in Fig.~\ref{fig3:triangular}). 
However, when $\lambda$ becomes large relative to the rate $\mu$ 
($\lambda = 0.8$ in Fig.~\ref{fig3:triangular}), the population density of the
intermediate species $C$ in the (quasi-)steady state is reduced so much 
($c_s = \frac{\mu}{\lambda} \approx 0.01$ in the figure) that the strong 
stochastic population oscillations cause the evolution trajectories to touch 
the absorbing phase boundary; thus, the simulation eventually terminates at one
of the absorbing states on the lines $(a, 0, b)$ which are not apparent in the 
mean-field analysis.

Furthermore, as shown in Fig.~\ref{fig3:lambda}, the quantitive relationship 
between the population density $c_s$ for the intermediate species $C$ in the
active coexistence state and the predation rate $\lambda$ numerically agrees 
remarkably well with the mean-field prediction $c_s = \mu/\lambda$. 
In Fig.~\ref{fig3:density}, we plot the temporal evolution of the three 
species' population densities for runs with predation rate $\lambda = 0.1$; the
population densities in the (quasi-)steady state (averaged over 50 runs at 
$t = 3000$ mcs) are found to be $(0.4 \pm 0.01, 0.1 \pm 0.003, 0.4 \pm 0.01)$, 
again fully consistent with the predictions of the mean-field approximation.
We also numerically investigated the influence of quenched spatial disorder on 
the dynamical evolution of the system. 
However, in stark contrast with the robust fitness enhancement in the 
two-species LV model, in the hierarchical three-species model spatial 
variability in the predation rate $\lambda$ does not generate any noticeable 
effect on the dynamical evolution of the system.  
These observations are further elucidated by the spatial distribution snapshot 
on a two-dimensional lattice depicted in Fig.~\ref{fig3:snpsh}: 
All particles appear randomly scattered on the lattice in the (quasi-)steady 
state, and no correlated spatio-temporal structures such as either clusters or 
spirals become manifest.
Consequently quenched spatial disorder cannot beneficially affect any 
correlations as in the LV model.

Our Monte Carlo simulations demonstrate that the spatial three-species 
hierarchical food chain system, with both nearest-neighbor particle pair 
exchange and hopping processes present, is quantitatively well-described by the
mean-field approximation. 
In the following subsection, we thus explore the respective role of particle 
pair exchange vs. hopping processes.

\subsection{Particle pair exchange processes and pure hopping processes} 
\label{pairexchange}

\begin{figure*}[!t]
\label{Fig4}
\begin{center}
\subfloat[Particle pair exchange processes]{\label{fig4:snpshex}
\includegraphics[width=0.3\textwidth]{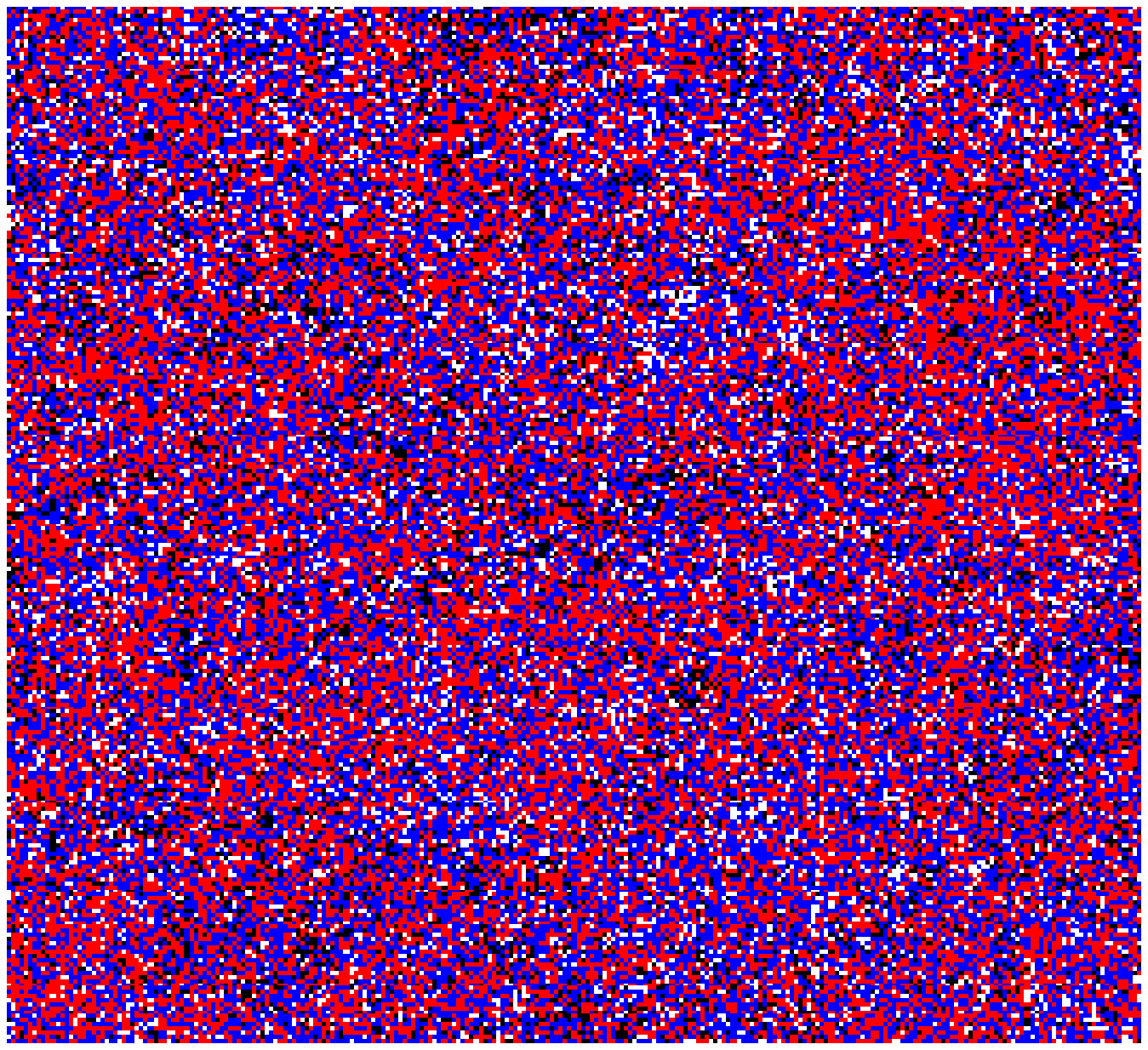}} 
\hspace{1.0in}\subfloat[Nearest-neighbor hopping]{\label{fig4:snpshhop}
\includegraphics[width=0.3\textwidth]{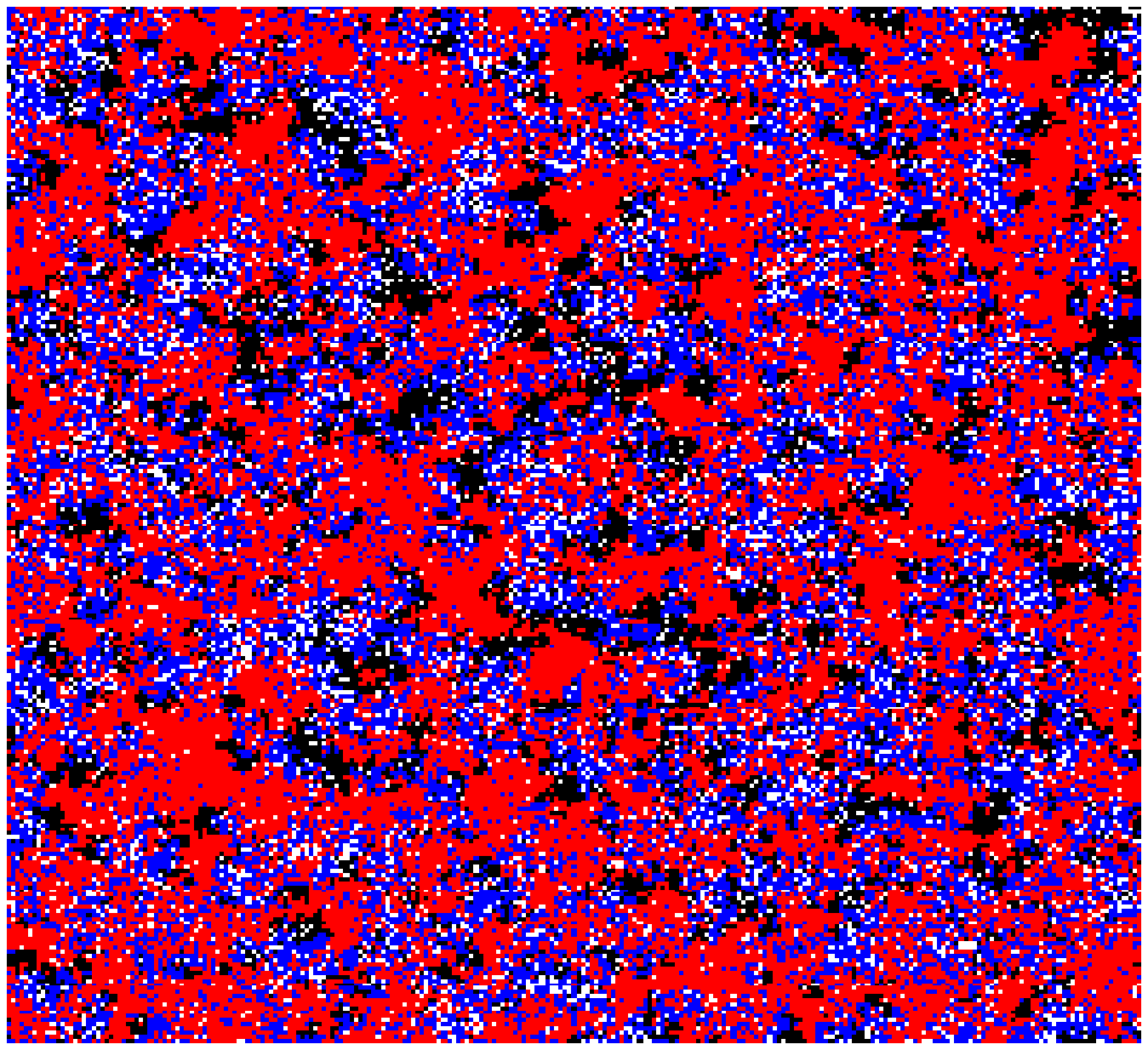}}
\\
\hspace{-0.0in}\subfloat[Temporal evolution for (a)]{\label{fig4:densityex}
\includegraphics[width=0.48\textwidth]{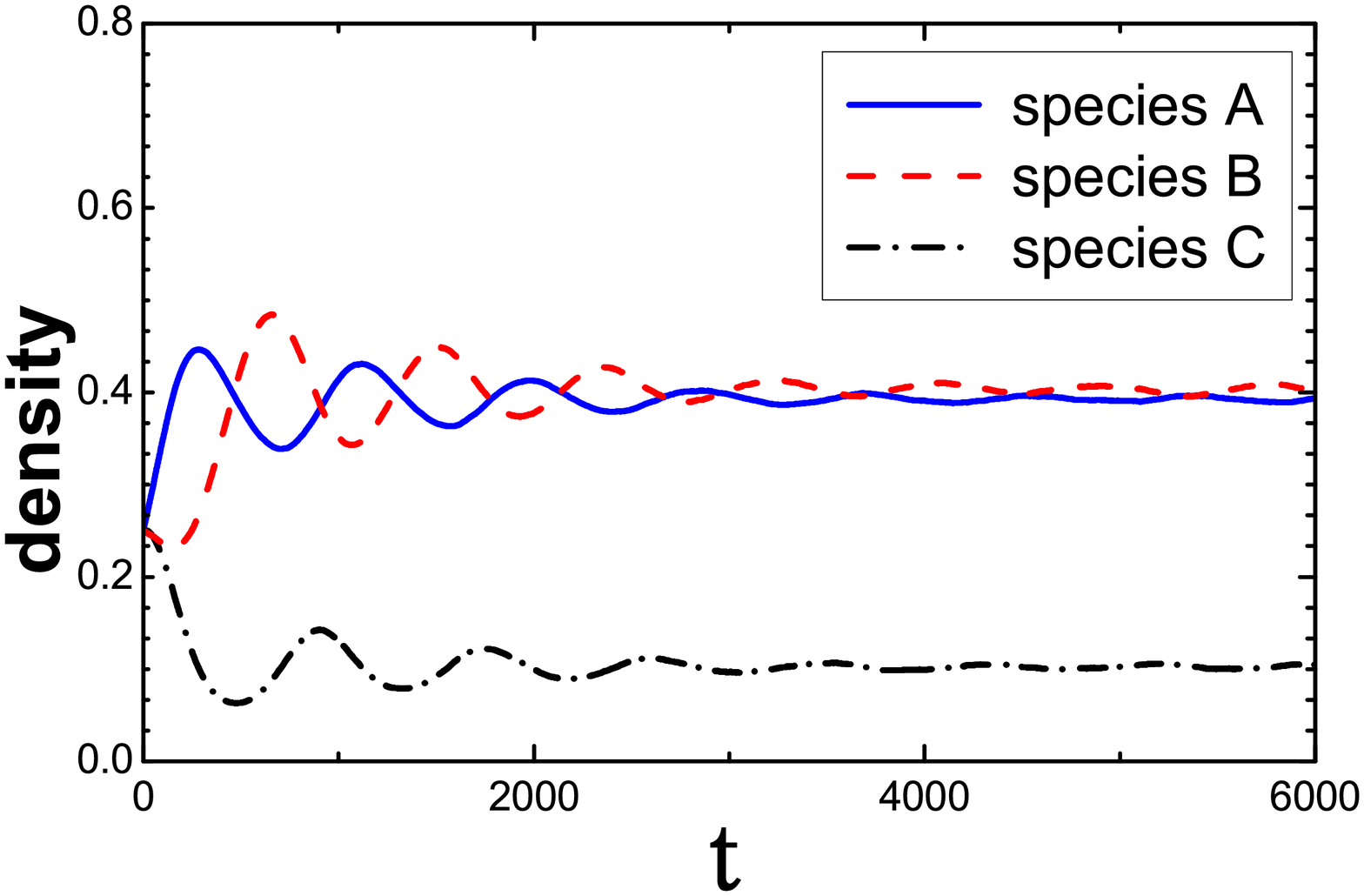}}
\hspace{0.0in}\subfloat[Temporal evolution for (b)]{\label{fig4:densityhop}
\includegraphics[width=0.48\textwidth]{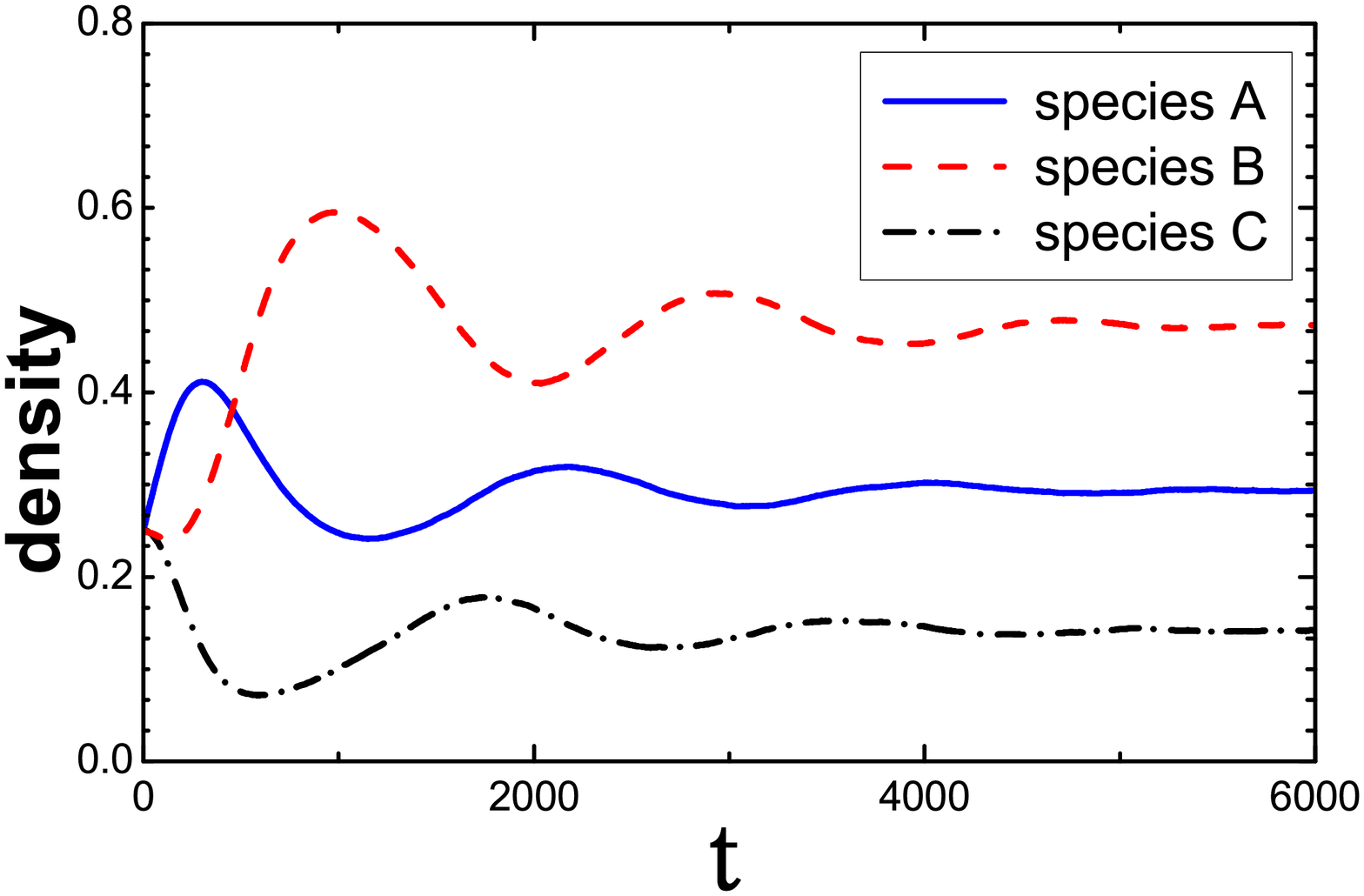}}  
\end{center}
\caption{{\it (Color online.)} (a) and (b): Snapshots of the spatial particle 
   distribution at $t = 6000$ mcs for a stochastic three-species hierarchical
   food chain system with $N = 256 \times 256$ sites, reaction rates 
   $\lambda = 0.1$, $\mu = 0.01$, and $\sigma = 0.1$, starting with equal 
   initial densities $a(0) = b(0) = c(0) = 0.25$. 
   Blue/dark gray: $A$, red/gray: $B$, black: $C$, white: empty.
   (c) and (d): Temporal evolution of population densities for simulation runs
   with (a) only nearest-neighbor particle exchange, and (b) only 
   nearest-neighbor hopping, with $N = 100 \times 100$ sites, averaged over 50
   simulation runs.}
\end{figure*}

In the above subsection, we have demonstrated that population spreading through
both nearest-neighbor particle pair exchange and hopping processes can well mix
the two-dimensional spatially extended three-species hierarchical food chain 
system.  
In order to further explore the origin for the appearance of mean-field 
behavior, we ran simulations wherein either the particle hopping or pair 
exchange processes were removed; Figs.~\ref{fig4:snpshex} and 
\ref{fig4:snpshhop} show snapshots of the resulting (quasi-)stationary particle
distributions.
In Fig.~\ref{fig4:snpshex} where only nearest-neighbor particle pair exchange 
was allowed, it is seen that the individuals for each species are uniformly
distributed on the lattice, and no particular spatial patterns emerge in the 
system.  
We depict the associated temporal evolution (up to $6000$ mcs) of the three
population densities for a smaller system ($N = 100 \times 100$) in 
Fig.~\ref{fig4:densityex}.
The population densities $(0.39 \pm 0.01,  0.10 \pm 0.01, 0.40 \pm 0.02)$ in 
the (quasi-) steady state (averaged over 50 runs at $t = 6000$ mcs) are again 
quantitatively consistent with the mean-field predictions.
Also, as one would expect, we have not observed any remarkable influence of 
quenched spatial disorder on the evolutionary dynamics in these systems.   
  
Next, we investigate systems wherein particle spreading occurs solely through 
nearest-neighbor hopping, see Figs.~\ref{fig4:snpshhop} and 
\ref{fig4:densityhop}.
In Fig.~\ref{fig4:snpshhop}, one notices distinct species clusters, as also 
observed in two-species LV systems; however, it is worthwhile noting that 
interactions between predators $A$ and prey $B$ do not directly take place at 
their respective cluster boundaries, but can only be indirectly realized via
interspersed clusters of the intermediate species $C$ (visible as black patches
in Fig.~\ref{fig4:snpshhop}). 
The required presence of the intermediate species clusters significantly 
coarse-grains the indirect reactions between predators $A$ and prey $B$, and 
effectively averages over spatially correlated structures.  
Again, we find that quenched spatial disorder does not noticeably affect the 
evolutionary dynamics of the system.  
However, as shown in Fig.~\ref{fig4:densityhop}, the population densities 
$(0.29 \pm 0.01, 0.14 \pm 0.01, 0.47 \pm 0.02)$ in the (quasi-) steady state  
(averaged over 50 runs at $t = 6000$ mcs) do not coincide with the predictions 
from the mean-field rate equations, i.e., nearest-neighbor hopping processes 
alone are not strong enough to effectively mix the two-dimensional 
three-species hierarchical food chain system.
We have furthermore checked that these numerical observations, here obtained 
for $\lambda = 0.1$, about the effect of either particle pair exchange 
processes or pure hopping processes also apply to systems with much 
smaller predation rate $\lambda = 0.02$.

In summary, we have demonstrated that the hierarchical three-species food chain
model with site restrictions and with particle spreading through pair exchange 
processes effectively washes out the formation of any spatial structures, hence
becomes well mixed, and behaves as in the mean-field predictions. 
As is also observed in the two-species LV system with site restrictions, the 
hierarchical food chain model also possesses a critical threshold for the
predation rate, where a phase transition from an inactive absorbing state to an
active coexistence state occurs.  
However, if particle spreading happens solely through nearest-neighbor hopping,
distinct species clusters emerge. 
Yet since intermediate species $C$ clusters are necessary to indirectly promote
predation interactions between the $A$ and $B$ species, the effective reaction
boundaries between the predators and prey become blurred, such that any spatial
variability still cannot remarkably influence the dynamical evolution of these
food chain systems.  
Consequently, at least for the parameter values and system sizes studied here,
the ensuing hierarchical three-species food chain model does not closely 
resemble the features of the corresponding two-species spatial stochastic LV 
system that is governed by strong fluctuations and marked correlations in the 
active coexistence state.

\section{Conclusion}
\label{conclu}

In this paper, we studied the connection between two three-species 
predator-prey systems and the well-understood two-species Lotka--Volterra 
model.
First, we explored the evolutionary behavior of the two minority species in the
``corner'' three-species cyclic rock--paper--scissors (RPS) model with strongly
asymmetric rates. 
We analytically demonstrate that in the mean-field limit the evolutionary 
dynamics of both minority species in the ``corner'' RPS model can be well 
approximated by the two-species LV model. 
Employing Monte Carlo simulation for a two-dimensional spatially extended
``corner'' RPS model, we found the population densities for those two minority 
species in the (quasi-)steady state to coincide with both the analytic 
predictions from mean-field approximation in the resulting two-species LV model
and its associated numerical simulation results.  
Furthermore, introducing quenched spatial disorder into the predation rate 
$\lambda$ in the ``corner'' RPS model, we observe that the influence of 
quenched spatial disorder on the evolution of system cannot be ignored, in
contrast to RPS systems far away from the ``corner'' of configuration space, 
wherein quenched spatial disorder has only minor effects on the evolution of 
the system \cite{Qian}:
The minority species' fitness in the ``corner''  RPS system is markedly 
enhanced due to the spatial variability in the predation rates, as is indicated
by larger population densities for both minority species in the (quasi-)steady 
state, and accompanied by shorter relaxation times and stronger localization of
the clusters for those two minority species.  
That is, the two minority species in the ``corner'' RPS model behave just like 
the predator and prey in the spatially extended two-species LV system.

We also investigated hierarchical three-species ``food chain'' systems, in 
which an additional intermediate speci\-es is inserted between the predators 
and prey in the classic two-species LV model. 
As observed in the spatial version of the two-species LV model with site 
occupancy restrictions, in the similarly restricted three-species ``food 
chain'' model with finite total carrying capacity, there exists a critical 
predation rate threshold, which in the thermodynamic limit represents a 
non-equilibrium phase transition from an inactive absorbing to an active 
coexistence state.
Our simulation results show that nearest-neighbor particle pair exchange 
processes are sufficient to wash out the formation of species clusters and
consequently generate a well-mixed spatial system, whose features can be 
quantitatively well approximated by the mean-field rate equations, and wherein
the influence of quenched spatial disorder may safely be ignored.  
Moreover, if we allow only nearest-neighbor hopping processes, and eliminate 
any pair exchanges, distinct species clusters appear in the coexistence state.
However, due to the interruption by intermediate species clusters, the 
predation processes between predator and prey are effectively coarse-grained,
and fluctuation and correlation effects on the resulting two-species dynamics
suppressed.
Consequently, quenched spatial disorder still cannot enhance the fitness of 
either species.
That is, as opposed to the ``corner'' rock-paper-scissors model with strongly 
asymmetric rates, the spatially extended hierarchical three-species ``food 
chain'' system actually does {\em not} behave like the two-species LV model. 
\\ 

\noindent
This work is in part supported by Virginia Tech's Institute for Critical 
Technology and Applied Science (ICTAS) through a Doctoral Scholarship, and the 
US National Science Foundation through grant No. DMR-1005417.
We gratefully acknowledge inspiring discussions with Uli Dobramysl and 
Michel Pleimling.

\end{document}